# Instrumental Tip-of-the-iceberg Effects on the Prompt Emission of Swift/BAT Gamma-ray Bursts

Michael Moss[1,2], Amy Lien[2,3,4,5], Sylvain Guiriec[1,2], S. Bradley Cenko[2,6], and Takanori Sakamoto[7]
[1] The Department of Physics, The George Washington University, 725 21st NW, Washington, DC 20052, USA; mikejmoss3@gmail.com
[2] Astrophysics Science Division, NASA Goddard Space Flight Center, Greenbelt, MD 20771, USA
[3] Center for Research and Exploration in Space Science and Technology (CRESST), Greenbelt, MD 20771, USA
[4] Department of Physics, University of Maryland, Baltimore County, 1000 Hilltop Circle, Baltimore, MD 21250, USA
[5] University of Tampa, Department of Chemistry, Biochemistry, and Physics, 401 W. Kennedy Blvd, Tampa, FL 33606, USA
[6] Joint Space-Science Institute, University of Maryland, College Park, MD 20742, USA
[7] College of Science and Engineering, Department of Physical Sciences, Aoyama Gakuin University, 5-10-1 Fuchinobe, Chuo-ku, Sagamihara-shi Kanagawa 252-5258, Japan

## Abstract

The observed durations of prompt gamma-ray emission from gamma-ray bursts (GRBs) are often used to infer the progenitors and energetics of the sources. Inaccurate duration measurements will have a significant impact on constraining the processes powering the bursts. The "tip-of-the-iceberg" effect describes how the observed signal is lost into background noise; lower instrument sensitivity leads to higher measurement bias. In this study, we investigate how observing conditions, such as the number of enabled detectors, background level, and incident angle of the source relative to the detector plane, affect the measured duration of GRB prompt emission observed with the Burst Alert Telescope on board the Neil Gehrels Swift Observatory (Swift/BAT). We generate "simple-pulse" light curves from an analytical fast rise exponential decay function and from a sample of eight real GRB light curves. We fold these through the Swift/BAT instrument response function to simulate light curves Swift/BAT would have observed for specific observing conditions. We find duration measurements are highly sensitive to observing conditions and the incident angle of the source has the highest impact on measurement bias. In most cases duration measurements of synthetic light curves are significantly shorter than the true burst duration. For the majority of our sample, the percentage of duration measurements consistent with the true duration is as low as ∼25%–45%. In this article, we provide quantification of the tip-of-the-iceberg effect on GRB light curves due to Swift/BAT instrumental effects for several unique light curves.

*Unified Astronomy Thesaurus concepts:* Gamma-ray bursts (629)

*Supporting material:* figure set

## 1. Introduction

Gamma-ray bursts (GRBs) are the most powerful explosions in the universe, ejecting a fraction of the Sun's rest mass energy in a matter of seconds. The most popular framework describing the GRB phenomenon assumes an accretion disk is formed around a compact object—either a rapidly spinning magnetar or a stellar-mass black hole—which has been recently produced by the core collapse of a massive star (Colgate 1968; Woosley et al. 1993; Galama et al. 1998; Hjorth et al. 2003) or the coalescence of two compact objects (Abbott et al. 2017; Eichler et al. 1989; Narayan et al. 1992; Tanvir et al. 2013). The accretion disk flows onto the compact object and powers a bipolar jet, accelerating particles (e.g., electrons, positrons, baryons) to highly relativistic speeds (Cavallo & Rees 1978; Rees & Meszaros 1992). As the highly energetic particles cool within the jet, gamma rays are produced via emission mechanisms such as synchrotron, Comptonization, inverse Compton, and photospheric processes, among others; this is called the prompt-emission phase of GRBs (Rees & Meszaros 1994; Piran 1999, 2004). The relativistic flow continues to propagate until the jet sweeps up enough circumburst and interstellar medium to quickly decelerate in a relativistic collisionless shock; this is the afterglow phase (Sari et al. 1998; Wijers & Galama 1999). Prompt emission is characterized by a bright flare of gamma rays that may last anywhere between a fraction of a second to hundreds of seconds and is mostly observed in the ∼kiloelectronvolt to ∼megaelectronvolt range, but can sometimes be observed down to optical wavelengths and up to gigaelectronvolt gamma rays (Akerlof et al. 1999; Hurley et al. 1994). The afterglow phase can be observed for hours to, occasionally, hundreds of days across the entire electromagnetic spectrum, down to radio frequencies and up to gigaelectronvolt gamma rays (Costa et al. 1997; Frail et al. 1997; van Paradijs et al. 1997; Atkins et al. 2000).

The duration of GRB prompt emission is traditionally measured using the $T_{90}$ algorithm, the duration which encompasses 5%–95% of the observed gamma-ray photon fluence within the 50–300 keV band in the observer frame (Kouveliotou et al. 1993). There is an observed bimodality in the $T_{90}$ distribution separated into short GRBs (SGRBs, $T_{90} \lesssim 2$ s) and long GRBs (LGRBs, $T_{90} \gtrsim 2$ s) (Kouveliotou et al. 1993). Coincidental detections between compact-binary mergers and SGRBs (Abbott et al. 2017; Tanvir et al. 2013) and between stripped-envelope core-collapse supernovae and LGRBs (Colgate 1968; Galama et al. 1998; Hjorth et al. 2003) provide a natural explanation for the observed dichotomy. Therefore, it is common to use $T_{90}$ measurements to infer GRB progenitor systems (e.g., either a collapsar or a compact-binary merger, $T_{90} \gtrsim 2$ s and $T_{90} \lesssim 2$ s, respectively).

However, the $T_{90}$ is not the intrinsic duration of a GRB: instrumental effects and observing conditions lower the signal-to-noise ratio (S/N) of an observed GRB, leading to $T_{90}$ measurements shorter than the true duration (Hakkila et al. 2000). These effects are instrument dependent (e.g., instrument energy bandpass, time resolution) and lead to differences in the $T_{90}$ distribution made with observations from different telescopes (see Figure 7 of Lien et al. 2016; Paciesas et al. 1999; Sakamoto et al. 2011; Bromberg et al. 2012; Svinkin et al. 2016; Tsvetkova et al. 2017; von Kienlin et al. 2020). The percentage of GRBs measured to be SGRBs by the Compton Gamma-Ray Observatory Burst And Transient Source Experiment (CGRO/BATSE), Fermi Gamma-Ray Space Telescope Gamma-ray Burst Monitor (Fermi/GBM), and Swift/Burst Alert Telescope (BAT) all differ from one another (i.e., ∼26%, ∼17%, and ∼10%, respectively; Paciesas et al. 1999; Lien et al. 2016; von Kienlin et al. 2020). It has even been argued that the separation time between SGRBs and LGRBs should vary depending on the instrument used (Bromberg et al. 2012). Furthermore, $T_{90}$ distributions do not typically include measurement uncertainties, which have a significant influence on the shape of the distributions and the separation of the SGRB and LGRB populations. A more robust method to determine the progenitor system is to use both spectral information and the $T_{90}$ measurement of a GRB, as SGRBs are typically observed with harder spectra than LGRBs (Dezalay et al. 1992; Kouveliotou et al. 1993). There have been proposals to establish a new classification system for GRBs by separating them into Type I and Type II GRBs (e.g., GRBs with compact-merger progenitors and those with supernova progenitors, respectively), which separate themselves when considering specific combinations of their observables (Gehrels et al. 2006; Zhang 2006; Lü et al. 2010; Minaev & Pozanenko 2020).

The trigger methods of an instrument also impact observed $T_{90}$ distributions. Frontera et al. (2009) found a discrepancy between the number of SGRBs detected with the Gamma-Ray Burst Monitor (GRBM) aboard BeppoSax compared to the number detected by CGRO/BATSE. For the majority of the mission, the BeppoSax/GRBM trigger system used a 1 s integration time to search for SGRBs, leading to a low triggering efficiency for SGRBs (Frontera et al. 2009).

Littlejohns et al. (2013) simulated a sample of low-redshift GRBs observed by Swift/BAT at higher redshifts to witness the evolution of their measured durations. The authors found that while time dilation always lengthens light-curve duration, measured durations are strongly impacted by the loss of burst structure into background noise as the S/N decreases. It was also shown that the relation between measured duration and redshift is highly dependent on the structure of the burst.

Measurements of GRB durations are sensitive to the observing conditions and the distance to the object. For instance, Kocevski & Petrosian (2013) investigated how accurate the $T_{90}$ measured by CGRO/BATSE was as a proxy for the intrinsic duration of a GRB. To do so, they simulated GRBs with fast rise exponential decay (FRED) light curves and folded them through BATSE instrument response functions. They found that the lack of very high $T_{90}$ values due to time dilation for high-redshift GRBs may be explained by low S/N; in some cases as much as 90% of the emission is buried into the noise, leading to $T_{90}$ measurements that significantly underestimate the intrinsic durations of the bursts in the observer frame.

It is recommended to use both spectral and temporal information to predict GRB progenitor systems, yet this does not always result in a clear answer (Kouveliotou et al. 1993). Only by directly observing a signal from the progenitor system (i.e., the supernova from a collapsar or the gravitational-wave signal from a compact-binary merger) can the type of progenitor be confirmed. GRB200826A is a SGRB characterized by a sharp pulse ($T_{90} \sim 1.1$ s, $z = 0.748$) and no evidence of any dim, long-lasting emission; this would infer a compact-binary merger progenitor (Mangan et al. 2020). However, the spectral behavior and energetics of the event were consistent with LGRBs from collapsar progenitors (Zhang et al. 2021). Additionally, the host galaxy is a low-mass, star-forming galaxy, typical for LGRBs (Rossi et al. 2021). Follow-up observations reveal excess emission in the afterglow light curve that cannot be explained as kilonova emission, but is consistent with supernova emission, which confirms that the progenitor system was a collapsar (Ahumada et al. 2021). Lü et al. (2014) have suggested that the "amplitude" of the GRB prompt emission, defined as the ratio between the measured peak flux and the flux background, is a necessary criterion to include to classify GRBs in order to distinguish between long/soft and short/hard GRBs. They also find that most SGRBs are likely not the "tip-of-the-iceberg" of LGRBs, but also show that most LGRBs would appear as rest-frame SGRBs above a certain redshift.

In this study we investigate the accuracy of $T_{90}$ measurements made by Swift/BAT as estimates of the intrinsic GRB prompt-emission durations. In Section 2, we describe Swift/BAT and introduce the instrument parameters which most significantly influence S/N. In Section 3, we describe the GRB sample used for our analysis. In Section 4, we outline our analysis methods. In Section 5, we discuss our results and their implications on the Swift/BAT GRB population.

## 2. The Neil Gehrels Swift Observatory and the Burst Alert Telescope

Swift was launched on 2004 November 20. Swift is a space-based multiwavelength observatory comprised of three instruments: a wide-field gamma-ray detector (BAT), a narrow-field X-Ray Telescope (XRT), and a narrow-field Ultra-Violet/Optical Telescope (UVOT; Gehrels et al. 2004; Barthelmy et al. 2005; Burrows et al. 2005; Roming et al. 2005).

Swift/BAT is a coded-mask aperture instrument with a 1.4 steradian field of view when half-coded (i.e., ∼10% of the sky) and is able to localize sources to within a few arcminutes (Barthelmy et al. 2005). The coded-mask aperture allows for imaging of photons between 15 and 150 keV, but the telescope is able to observe noncoded photons up to ∼350 keV. In this work we investigate how the observing conditions of Swift/BAT influence the accuracy of $T_{90}$ measurements. Namely, we investigate the effects due from changes in the active number of enabled detectors, the angle of the source from the detector bore sight, and average background level. These three parameters were found to have the strongest impact on duration measurements made by Swift/BAT; each parameter is described below.

### 2.1. Detector Plane versus Time

At the start of the mission, Swift/BAT had 32,768 CdZnTe detectors (each $4 \times 4 \times 2$ mm); however, as the instrument has aged, some detectors have become permanently noisy and are therefore turned off; this results in the number of enabled detectors steadily decreasing (see Figure 1(a)). In 2019, the yearly averaged number of enabled detectors (NDETS) was

∼16,000. In 2020, due to a series of controlled reboots performed on the BAT instrument, the yearly averaged NDETS increased to ∼18,000. Although we display the yearly averaged NDETS in Figure 1(a), some observations have reached NDETS levels as low as ∼10,000. As the number of enabled detectors reduces, so too does the instrument sensitivity (see Figure 4 in Lien et al. 2014).

### 2.2. Incident Angle from Detector Bore Sight

The coded-aperture mask technique is useful in X-ray and gamma-ray astronomy because it allows an instrument to have a large field of view while maintaining imaging capabilities for improved source localization (Barthelmy et al. 2005); however, the sensitivity of coded-mask instruments quickly decreases as the incident angle of the source with respect to the telescope bore sight, $\theta_{\rm inc}$, increases (see Figure 2 in Lien et al. 2014). The coded-aperture method defines the partial coding fraction (PCODE) as the fraction of the detector illuminated by a source through the mask. It is correlated to the source incident angle, but it also accounts for instrument geometry; PCODE = 1 corresponds to an on-axis observation and a fully illuminated detector plane, PCODE = 0 indicates an incident angle of $\gtrsim 70°$. The PCODE accounts for the noncircularity of the Swift/BAT field-of-view contours on the detector plane. The contours are not rotationally symmetric around the detector bore sight due to the geometry of the detector (Barthelmy et al. 2005). The PCODE reflects the effective area of the Swift/BAT detector plane in use during an observation.

The distributions of PCODE and $\theta_{\rm inc}$ are displayed in Figure 2(a). The distributions are created from the PCODE and $\theta_{\rm inc}$ values reported for all GRBs in the Swift/BAT GRB catalog data tables. The PCODE and $\theta_{\rm inc}$ reported are the values taken at the time of the GRB trigger. We can see that the PCODE distribution is somewhat uniformly distributed, but has a peak at PCODE = 1. The $\theta_{\rm inc}$ distribution is a combination of the cosine effect from projecting a three-dimensional sphere onto a two-dimensional plane, and the decrease in sensitivity at larger incident angles. In Figure 2(b) the PCODE and $\theta_{\rm inc}$ values for all Swift/BAT GRBs are shown. From this plot we can see how the $\theta_{\rm inc}$ distribution can spread out into the PCODE distribution we observe. A PCODE = 1 can be obtained even at $\theta_{\rm inc} \sim 20°$ depending on the $\phi$ angle of the source relative to the detector plane.

### 2.3. Average Background Level

Swift/BAT is a background-dominated instrument (Markwardt et al. 2007). In our simulations we assume flat backgrounds. From a sample of 1350 observations of GRBs made with Swift/BAT, the average background levels were calculated using the background emission that occurred within the $T0$–110 s to $T0$–10 s interval. We show the evolution of the average background level by separating the background sample into four distributions with respect to the date of the observations (date format: mm/yyyy). The average total background level decreases over time due to the decreasing number of enabled detectors over time (see Figure 1(b)). When the background is normalized by the number of enabled detectors turned on during an observation, the counts s$^{-1}$ detector$^{-1}$ has remained the same since the launch of Swift (see Figure 1(c)).

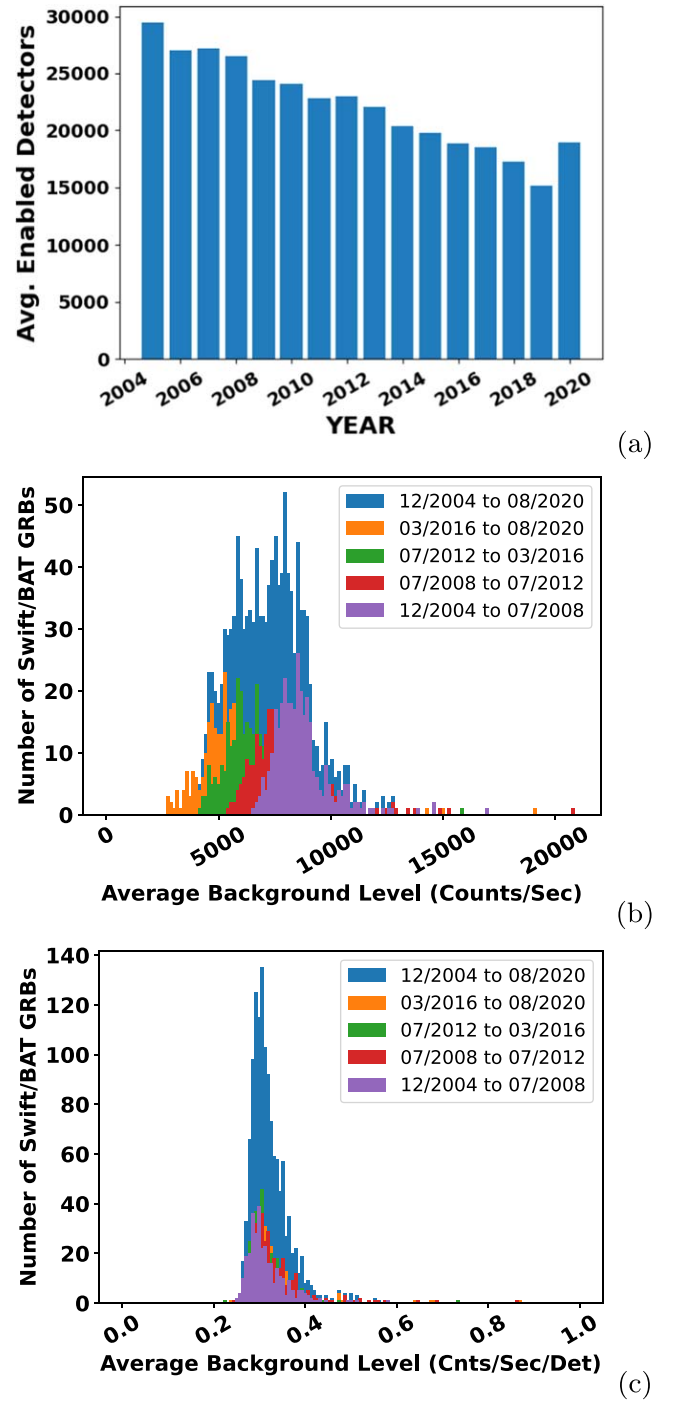

**Figure 1.** (a) The yearly averaged number of enabled detectors on board Swift/BAT is steadily decreasing as detectors become permanently noisy and are ultimately turned off. In 2004, BAT had 32,768 enabled detectors and, in 2020, the yearly averaged NDETS was ∼18,000, but has been as low as ∼10,000 during some observations. (b) Displayed is a distribution of 1350 average background levels measured during the $T0$–110 s to $T0$–10 s interval before Swift/BAT observations of GRBs (in blue). The 1350 measurements are separated into four time intervals to show the evolution of the background over time (date format: mm/yyyy). (c) Similar to the figure in panel (b), but normalized by the number of enabled detectors in use during the observation of each source. It is clear the the average background of Swift/BAT has not changed and the decrease seen in panel (b) is an effect of the decreasing number of detectors.

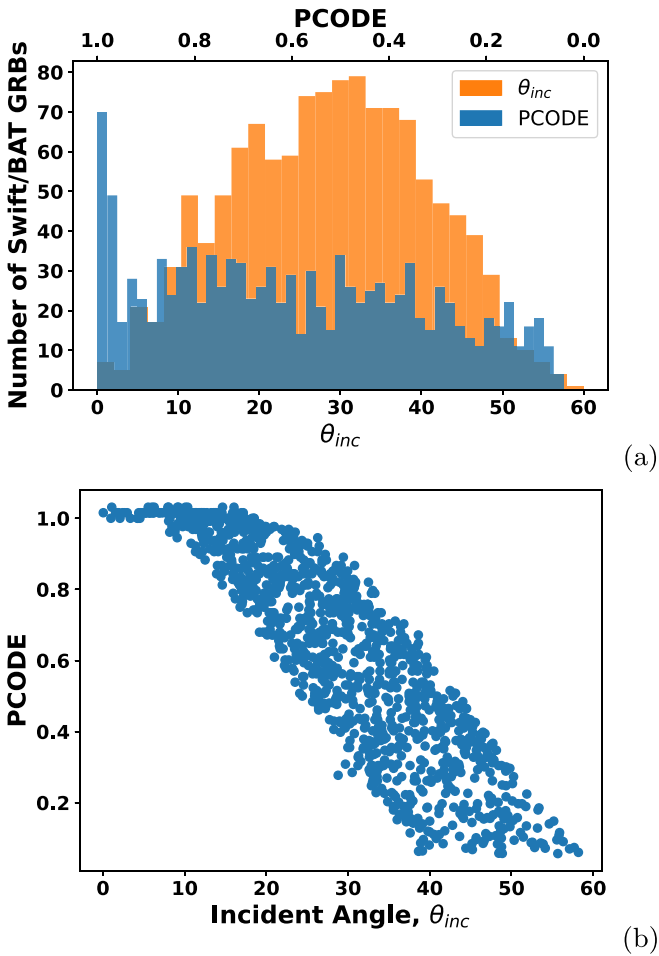

Figure 2. (a) Shown are the distributions of PCODE (blue) and source incident angle (orange) for the entire Swift/BAT GRB catalog. The distributions comprise the PCODE and $\theta_{\rm inc}$ measured at the time of the GRB trigger. (b) Displayed are the PCODE and $\theta_{\rm inc}$ values at the time of trigger for all Swift/BAT GRBs. As we can see, $\theta_{\rm inc}$ and PCODE are related, but the relation is not simple due to the fact PCODE accounts for detector geometry.

### 2.4. Trigger Methods

Swift/BAT has two trigger methods: rate trigger and image trigger (Barthelmy et al. 2005). The rate-trigger method uses over 600 different photon-rate criteria to search for a significant signal above background. Each criterion specifies a unique combination of S/N threshold, foreground/background intervals, elapsed time between the intervals, and one of four Swift/BAT energy bands to search within (i.e., 15–25 keV, 15–50 keV, 25–100 keV, and 50–350 keV). The rate-trigger method requires both an image trigger and one of the rate criteria to be met, while an image trigger only requires an image-trigger threshold to be met. The image-trigger method uses images created with various integration times (all ⩾1 minute) to identify sources within the field of view and only requires that one of the image-threshold criteria are met. We adopted the trigger simulator developed in Lien et al. (2014), which mimics the onboard trigger algorithms.

### 3. Light-curve Sample

In order to investigate the effects that observing conditions have on GRB prompt-emission duration, we will apply a duration-measurement algorithm similar to the one employed by the standard Swift/BAT pipeline to a sample of light curves.

Table 1
Sample of Synthetic FRED Light Curves (Hakkila & Preece 2014)

| Label | $A$ (counts s$^{-1}$) | $t_s$ (s) | $\tau_1$ | $\tau_2$ | Fluence (erg cm$^{-2}$) | $T_{90}$ (s) |
|---|---|---|---|---|---|---|
| (1) | (2) | (3) | (4) | (5) | (6) | (7) |
| FRED1 | 20000 | −0.1 | 1 | 4.5 | $1.44 \times 10^{-5}$ | 14.14 |
| FRED2 | 9000 | −0.1 | 1 | 4.5 | $6.07 \times 10^{-6}$ | 14.13 |
| FRED3 | 5000 | −0.1 | 1 | 4.5 | $3.64 \times 10^{-6}$ | 14.17 |
| FRED4 | 2500 | −0.1 | 1 | 4.5 | $1.54 \times 10^{-6}$ | 14.14 |

**Note.** $A$ is the pulse amplitude, $t_s$ is the pulse start time, $\tau_1$ and $\tau_2$ characterize the pulse rise and decay. The total fluence and $T_{90}$ of the synthetic light curves are reported in columns 6 and 7, respectively.

In this initial study, we focus on light curves with a "simple-pulse" shape, where "simple-pulse" light curves are designated as either "FRED-like" if the light curve has a rise and fall similar to a FRED light curve or as "symmetric-like" if the light curve rises and falls more gradually and symmetrically. It should be noted that our definition of simple-pulse structure is not and need not be rigorous, but is used as a loose guide line to select bursts with relatively simple structure as compared to the more complex bursts (e.g., with multiple bright peaks or multiple emission periods). Although we use simple-pulse light curves, we have selected light curves with a few unique features in order to investigate how light-curve phenomenology plays a role in the accuracy of the $T_{90}$. In the next sections we describe how we generated or selected and prepared the synthetic and real light-curve samples.

### 3.1. Synthetic FRED Light Curves

To create ideal benchmark cases for our results, we created four synthetic FRED light curves. We use the Hakkila & Preece (2014) FRED function form:

$$I(t) = A\lambda e^{[-\tau_1/(t-t_s)-(t-t_s)/\tau_2]}, \qquad (1)$$

where $A$ is the pulse amplitude, $t_s$ is the pulse start time, $\tau_1$ and $\tau_2$ characterize the pulse rise and decay, respectively, and $\lambda = e^{[2(\tau_1/\tau_2)^{1/2}]}$. The pulse peak time occurs at $t_{\rm peak} = t_s + \sqrt{\tau_1\tau_2}$ (Hakkila & Preece 2014). The parameters used to create our benchmark FRED light curves are displayed in Table 1 and the corresponding light curves are shown in Figure 3. We vary the count rate in each time bin according to a Gaussian centered on the theoretical count rate with a standard deviation equal to the square root of the count rate (i.e., $N_{\rm FRED}(t) = Norm(\mu = I(t), \sigma = \sqrt{I(t)})$). We analytically find the $T_{90}$ of each, where the $T_{90}$ is defined as the duration which encompasses 5%–95% of the total fluence (see the rightmost column in Table 1). For all four FREDs, we set the spectrum in the 15–350 keV band as a power law (PL) with a photon index of $\alpha = -1$, a typical photon index found in Swift/BAT GRBs (Lien et al. 2016). For our sample of FRED light curves, we keep all parameters the same except the amplitudes. The amplitude values are selected to produce FRED light curves with mask-weighted count rates similar to those found for typical Swift/BAT GRBs.

### 3.2. Real GRB Light Curves

We select a sample of eight real Swift/BAT-detected GRBs observed between 2009 and 2015. Simple-pulse burst structure was determined by visual inspection upon observation with Swift/BAT.[8] We briefly describe the shape of each burst in the rightmost column of Table 2 and display the corresponding light curves in Figure 9. GRB data were taken from the online Swift/BAT Gamma-Ray Burst Catalog.[9]

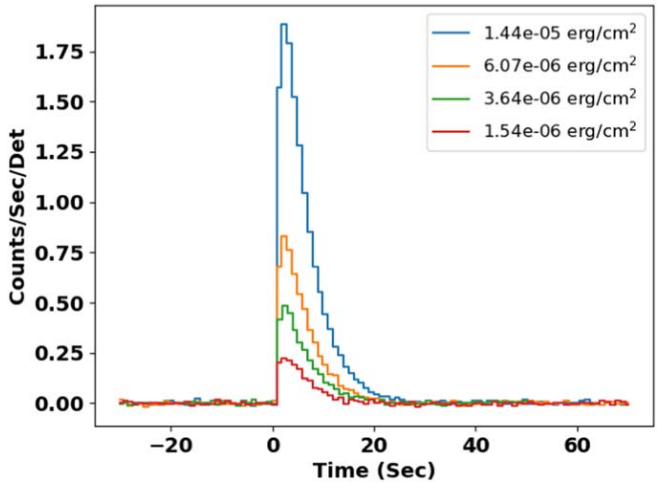

**Figure 3.** Four FRED-shaped light curves with different normalizations corresponding to evaluations of Equation (1) with parameters listed in Table 1.

Our simulations use the light curves of GRBs observed by Swift/BAT, meaning that they have already been affected by instrumental parameters; because of this our results should be considered self-consistent and not used to infer the intrinsic behavior for the GRBs they came from.

## 4. Simulation of Swift/BAT GRB Light Curves

For our analysis we define $T_{90}^{\rm BAT}$ ($T_{100}^{\rm BAT}$) as the duration which encompasses 5%–95% (0%–100%) of the light-curve background-subtracted count fluence in the 15–350 keV band in the observer frame (i.e., the Swift/BAT band). $T_{90}^{\rm BAT}$'s calculated for template light curves (i.e., before folding the light curve with the instrument response function) represent the true $T_{90}^{\rm BAT}$ of the light curve and are thus designated $T_{90,\rm true}^{\rm BAT}$. $T_{90}^{\rm BAT}$'s measured for simulated light curves (i.e., after folding with the instrument response function) are designated as $T_{90,\rm sim}^{\rm BAT}$.

We study how observing conditions affect the $T_{90}^{\rm BAT}$ of synthetic FRED light curves and real simple-pulse GRB light curves. To do so, we fold template light curves through Swift/BAT instrument response functions at varying observing conditions and use a Bayesian block method to measure their $T_{90,\rm sim}^{\rm BAT}$ values (Scargle 1998). A detailed description of our simulation method is given below. A simple schematic of the pipeline is shown in Figure 4.

We generate template light curves from Equation (1) and from our sample of real Swift/BAT GRBs. Swift/BAT light-curve counts are reported as mask-weighted count units, which are defined as "background subtracted counts per fully illuminated detector for an equivalent on-axis source"[10] (Markwardt et al. 2007). This means counts per detector per sec may be negative when close to the background level (top plot in Figure 5). For our light-curve templates we only select the positive counts of the background-subtracted and detector-corrected emission that occurred during the $T_{100}$ interval reported by the Swift/BAT team;[11] by doing so, we are assuming all the positive counts in a template light curve come from real burst emission. An example of this process used on the light curve of GRB050219A is shown in Figure 5. The template light curves (with units of photons det$^{-1}$ s$^{-1}$) are turned into template energy flux curves (with units erg det$^{-1}$ s$^{-1}$) by multiplying by the average energy per photon. The average energy per photon is found by taking the ratio of the integral of the spectrum across the 15–350 keV band and the total number of photons in the template light curve. This is necessary to convolve the template light curve with the instrument response function.

Following the procedure in Lien et al. (2014), we simulate the light curve Swift/BAT would observe for any given energy flux light curve, spectrum, and set of instrument parameters (e.g., angle of incidence, number of activated detectors, background level).[12] The best-fitting spectral function for most GRBs in the Swift/BAT GRB catalog is typically a PL but can occasionally be a cut-off power law (CPL) when the break energy is within or near the 15–350 keV band. Both of these spectral functions can be approximated by a Band function when considering the spectrum in a narrow energy band. In this work, we assume a Band function spectrum for all GRBs with the observed spectral index and, when observable, break energy (Band et al. 1993). For the spectrum of each burst we use the best-fit time-integrated spectral function and parameters reported in the BAT GRB catalog (Lien et al. 2016). One limitation of using the Band function is that the bolometric integration of the Band function evaluates to infinity if a low-energy PL index of $\leqslant -2$ is used. For this reason any burst used as a template in our pipeline with an observed PL index of $\alpha_{\rm PL} \leqslant -2$ has its index set to $\alpha_{\rm BAND} = -1.99$. We do not include any GRBs with $\alpha_{\rm PL} \leqslant -2$ in the sample used in this work.

We keep the spectrum constant throughout the duration of a GRB to reduce the computational resources required during simulations, meaning there is no time-evolution of the spectra, although the pipeline is able to use evolving spectra if desired. We have found that bursts with more complex structure than a FRED-like simple pulse (i.e., multipulsed or extended emission) are not well represented by a constant spectrum. For the simple-pulse sample used in this work, we find that using a constant spectrum produces results that do not differ significantly from results obtained using time-dependent spectra.

The Swift/BAT response function depends on the source incident angle and azimuthal angle in relation to the detector plane. For our simulations we split the detector plane into 31 regions, following the work of Lien et al. (2014; see their Figure 1). Each region is assigned a unique instrument response function that well represents the instrument response at that location on the detector plane. We label each region with a GridID number running from 1 to 33 (the grid spaces 06 and 28 are unused). The GridID serves as a proxy to specifying the

---

[8] A list of visually simple-pulsed bursts observed by Swift/BAT can be found at https://swift.gsfc.nasa.gov/results/batgrbcat/summary_cflux/summary_GRBlist/GRBlist_single_pulse_GRB.txt.
[9] https://swift.gsfc.nasa.gov/results/batgrbcat/
[10] https://swift.gsfc.nasa.gov/analysis/threads/batfluxunitsthread.html
[11] The total number of counts detected from a GRB in the observer frame is smaller than the number of photons emitted by the source in the source frame. We simply use the observed light curves as templates for realistic burst structure.
[12] The simulation code can be found here: https://userpages.umbc.edu/~alien/trigger_simulator/.

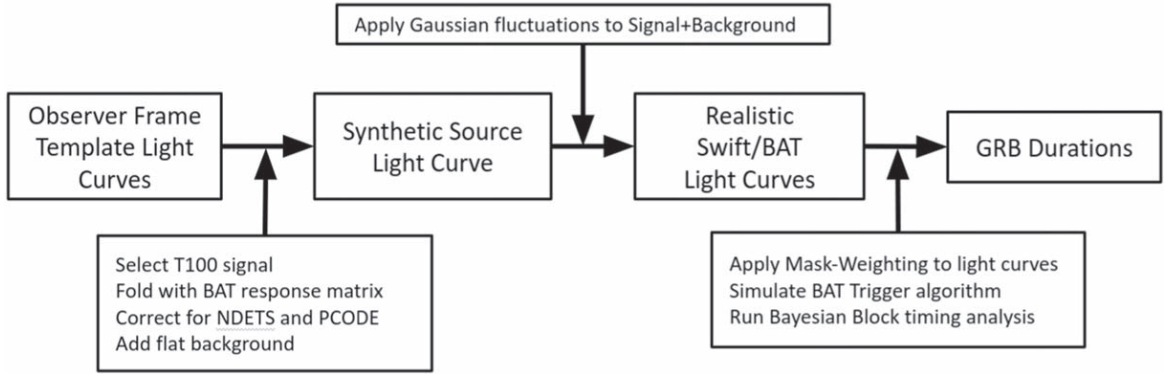

**Figure 4.** A schematic of our simulation pipeline used to generate synthetic $T_{90}$ measurements made with Swift/BAT. Synthetic FRED light curves and real Swift/BAT GRB light curves are used are to create template light curves. These templates are then folded with the instrument response matrix to create synthetic-source light curves. The PCODE or incident angle used during simulations is associated with a particular GridID, which in turn specifies which instrument response matrix to use (Lien et al. 2014). The count rates are then scaled by a factor to accurately reflect the instrument sensitivity at different numbers of enabled detectors (i.e., a factor of NDETS/32,768). A flat background is added and Gaussian fluctuations are applied to the synthetic light curves to simulate light curves that would have been observed by Swift/BAT. A Bayesian block method is used to measure the $T_{90}$ of the synthetic observed light curves.

Table 2
Sample of LGRBs with a "Simple-pulse" Shape.

| GRB Name | $z$ | $T_{90}$ (s) | Fluence (erg cm$^{-2}$)) | $\alpha$ | PCODE | $\theta_{\rm inc}$ | S/N | Description |
| --- | --- | --- | --- | --- | --- | --- | --- | --- |
| (1) | (2) | (3) | (4) | (5) | (6) | (7) | (8) | (9) |
| GRB160314A | 1.726 | 8.73 | $3.75 \times 10^{-07}$ | −1.53 | 0.75 | 19°.69 | 14.18 | Dim GRB |
| GRB150314A | 1.76 | 14.8 | $5.13 \times 10^{-05}$ | −1.08 | 0.344 | 35°.1 | 256 | FRED-like with dim tail |
| GRB120119A | 1.73 | 68.0 | $3.17 \times 10^{-05}$ | −1.38 | 1.02 | 5°.13 | 45.46 | Symmetric-like |
| GRB110422A | 1.77 | 25.8 | $5.56 \times 10^{-05}$ | −0.831 | 0.227 | 44°.7 | 27.95 | Symmetric-like |
| GRB090510 | 0.903 | 5.664 | $1.46 \times 10^{-06}$ | −1.06 | 0.162 | 46°.07 | 145.49 | Short, hard spike with soft tail |
| GRB071010B | 0.947 | 36.124 | $6.21 \times 10^{-06}$ | −1.97 | 0.8438 | 29°.04 | 52.96 | FRED-like with dim pretrigger emission |
| GRB051111 | 1.55 | 64.0 | $7.94 \times 10^{-06}$ | −1.32 | 0.594 | 27°.2 | 37.09 | Broad FRED-like |
| GRB050219A | 0.211 | 23.8 | $4.53 \times 10^{-06}$ | −0.124 | 0.232 | 43°.1 | 14.69 | Symmetric-like |

**Note.** All values reported in the table are from the online Swift/BAT Gamma-Ray Burst Catalog. The observatories and methods used to measure the redshift, $z$, for each burst are cited in the Swift/BAT GRB catalog. The $T_{90}$, fluence, and the spectral index, $\alpha$, are all measured by Swift/BAT. The PCODE describes the effective area of the Swift/BAT detector plane in use during the observation. The PCODE is related to the incident angle of the source from the detector bore sight, $\theta_{\rm inc}$ (see Figure 2). The observed rate S/N ratio (as opposed to image S/N) is given in the second-to-last column. A short description of the burst light curve is given in the final column. The light curves can be seen in Figure 9.

burst incident and azimuthal angles (and, hence, also a proxy of the PCODE). For each simulation we specify a PCODE to use and we then find the GridID with the nearest PCODE value. The selected GridID designates which response function to use for the simulation. GRB spectra are convolved with Swift/BAT response functions in order to calculate observed count rates in each Swift/BAT energy channel. The count rates are then scaled by a factor to accurately reflect the instrument sensitivity at different numbers of enabled detectors (i.e., a factor of NDETS/32,768).

By multiplying the template energy flux light curves by the average rates found from the synthetic spectra, we create synthetic-source light curves. To create synthetic observed light curves which represent light curves that Swift/BAT would have observed for a particular set of observing conditions, we add a flat background to the synthetic-source light curves and, following the procedure used for the FRED light curves, we fluctuate the counts in each time bin according to a Gaussian distribution centered around the expected number of counts in order to simulate statistical noise fluctuations (i.e., $N_{\rm obs}(t) = Norm(N_{\rm source}(t), \sqrt{N_{\rm source}(t)}))$.

Our simulations are evaluated using three NDETS values, three PCODE values, and three average background levels. The NDETS values we use are approximately the yearly average number of enabled detectors in 2004, 2013, and some particularly low levels during observations made in 2019 (i.e., NDETS = 30,000, 20,000, and 10,000, respectively). The PCODE values we select represent an on-axis observation, a typical off-axis observation, and an extreme off-axis observation (i.e., PCODE = 1, 0.522, and 0.101, corresponding to $\theta_{\rm incident} = 0°$, 44°.988, and 64°.286, respectively). The background levels we select represent a very low background, a typical background, and a high background for Swift/BAT (e.g., ∼1000 c s$^{-1}$, ∼9500 c s$^{-1}$, and ∼22,000 c s$^{-1}$, respectively; see Figure 1(b)).

We test whether Swift/BAT would trigger on these simulated observed light curves by approximating the true Swift/BAT trigger search following the procedure in Lien et al. (2014). To measure the duration of the simulated light curves, we use the FTOOLS function battblocks[13], which uses a similar Bayesian block-searching analysis as the pipeline used

---
[13] https://heasarc.gsfc.nasa.gov/lheasoft/ftools/caldb/help/battblocks.html

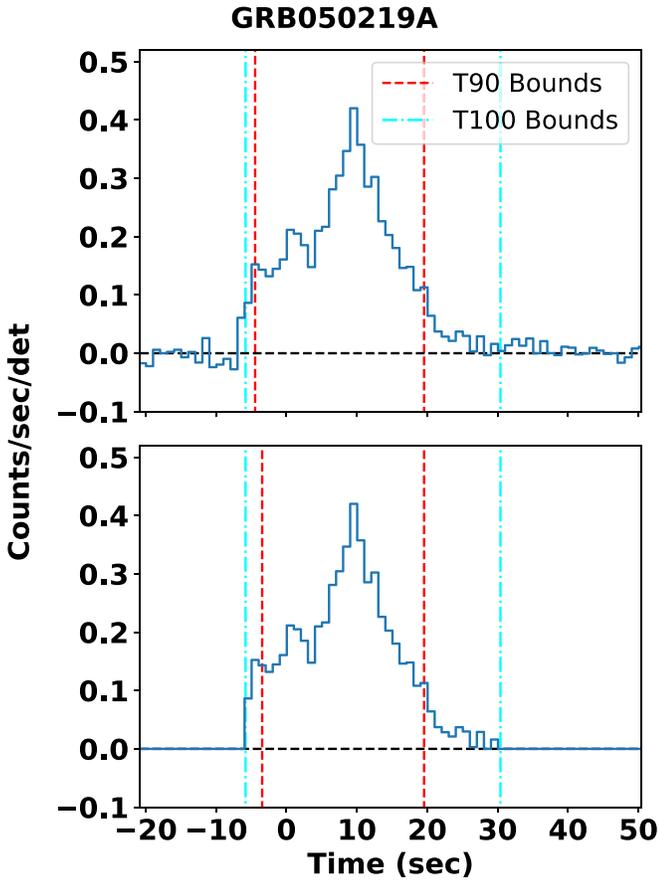

**Figure 5.** Top: the observed light curve of GRB050219A within the $T_{100}$ reported in the BAT GRB catalog (Lien et al. 2016). Bottom: the template light curve generated from the signal within the $T_{100}$ interval of the burst with all negative mask-weighted counts removed.

by the standard Swift/BAT ground analysis to determine burst durations. To correctly scale the S/N and the associated uncertainty in each time bin, mask weighting must be applied to the synthetic observed light curves. Proper mask weighting involves complex ray-tracing, as is done for official BAT products (Markwardt et al. 2007); we avoid this computationally intensive process by applying a simple and empirical mask-weighting approximation,

$$\text{MaskWeightedLC} = \frac{(\text{LC} - \text{Background})}{\cos(\theta_i) * \text{NDETS} * \text{PCODE} * f(\text{PCODE})}, \quad (2)$$

where "LC" is the count rate in each time bin of the synthetic observed light curve, "Background" is the average background of the light curve, BGDLVL, and $\theta_i$ is the angle of incidence. There is an additional light-curve efficiency factor that roughly depends on the PCODE of the observed burst: this factor is due to the fast Fourier transform convolution of the mask and the detector plane. To properly calculate the factor the Swift/BAT standard pipeline uses complex ray-tracing algorithms. In order to avoid the computationally intensive calculations, we calculated this correction factor for a sample of 100 GRBs and empirically fit the factors with a 2nd-order polynomial as a function of the PCODE; we include this as correction $f$(PCODE) (see Figure 6). The best-fit polynomial we found

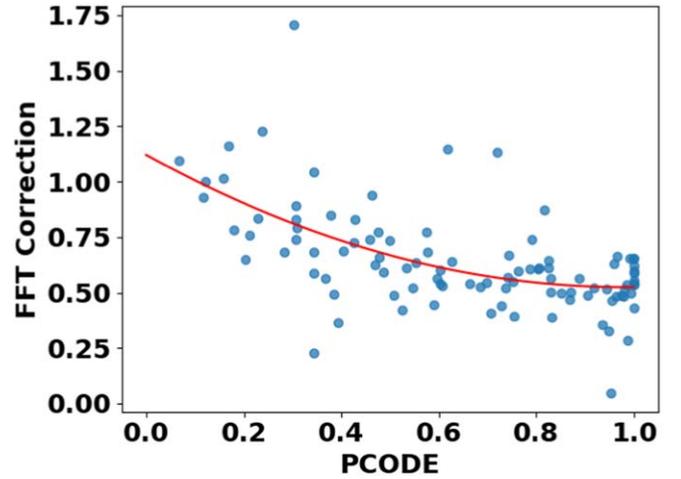

**Figure 6.** Fast Fourier transform correction factor for a sample of 100 GRBs observed by Swift/BAT and the best-fit 2nd-order polynomial (as given by Equation (3)).

is given by

$$f = 0.618 * \text{PCODE}^2 - 1.214 * \text{PCODE} + 1.121. \quad (3)$$

We then apply the `battblocks` algorithm to the mask-weighted light curves in order to obtain duration measurements for the synthetic observed light curves, $T_{90,\text{sim}}^{\text{BAT}}$. We do not account for high-energy photons penetrating through instrument structure. While this may have an influence on bursts with particularly hard spectra, this is a small effect in any case.

## 5. Implications for the Swift/BAT GRB Population

The results of the simulations outlined in Section 4 show that the $T_{90}$ of simple-pulse GRB prompt emission measured by Swift/BAT are highly impacted by observing conditions. As observing conditions become worse, (i) $T_{90}$ measurements become increasingly biased toward shorter durations, and (ii) the measurement uncertainties become larger.

We report the $T_{90,\text{sim}}^{\text{BAT}}$ distributions for several sets of observing conditions for all synthetic and real GRBs in our sample in Appendix B. As the S/N of a simulated light curve decreases the bias on the $T_{90,\text{sim}}^{\text{BAT}}$ increases toward shorter durations, yet there does not appear to be a quantifiable trend as it is highly entangled with the shape of the light curve (see Section 5.1 and Appendix D). Light curves simulated with typical observing parameters for Swift/BAT (e.g., PCODE = 1, 0.522; NDETS = 20,000; BGDLVL = 95,000 counts s$^{-1}$) were typically consistent in only ~25%–45% of simulations (see Table 3).

In the following sections we first discuss the measurement bias of $T_{90,\text{sim}}^{\text{BAT}}$ and the fluence as a function of S/N. We then discuss how low-intensity tails of GRB emission can be easily missed due to high background levels. Lastly, we discuss the possibility that some SGRBs observed by Swift/BAT have a chance of being collapsar events, which are typically associated with LGRBs.

### 5.1. Bias of $T_{90,\text{sim}}^{\text{BAT}}$ and Fluence Measurements as a Function of S/N

For analytical FRED light curves, the fraction of duration measurements made on simulated light curves which are consistent to within 1$\sigma$ of the analytical duration of the FRED

Table 3
Fraction of Simulated Light Curves for which the Bayesian Block Algorithm Was Able to Obtain Duration Measurements

| GRB Name | $f$(measurable) | $f$(consistent, 3$\sigma$) | $f$(consistent, 1$\sigma$) | $T_{90,\mathrm{true}}$ (s) | Ave. $T_{90,\mathrm{sim}}$ (s) | 90% CI (s) 68% CI (s) |
|---|---|---|---|---|---|---|
| (1) | (2) | (3) | (4) | (5) | (6) | (7) |
| FRED1 ($1.44 \times 10^{-5}$ erg cm$^{-2}$) | 0.914 | 0.826 | 0.271 | 14.14 | 11.62 | [4.008, 15.108] [6.008, 14.108] |
| FRED2 ($6.07 \times 10^{-6}$ erg cm$^{-2}$) | 0.787 | 0.828 | 0.329 | 14.13 | 10.86 | [5.008, 14.108] [6.008, 14.108] |
| FRED3 ($3.64 \times 10^{-6}$ erg cm$^{-2}$) | 0.709 | 0.618 | 0.259 | 14.17 | 10.22 | [4.008, 15.108] [5.008, 14.108] |
| FRED4 ($1.54 \times 10^{-6}$ erg cm$^{-2}$) | 0.571 | 0.468 | 0.224 | 14.14 | 8.64 | [4.008, 14.108] [5.008, 12.108] |
| GRB160314A | 0.289 | 0.440 | 0.346 | 8.64 | 6.95 | [0.942, 9.442] [2.042, 9.042] |
| GRB150314A | 0.990 | 0.120 | 0.069 | 16.0 | 11.14 | [4.003, 19.103] [6.003, 15.103] |
| GRB120119A | 0.911 | 0.182 | 0.074 | 83.0 | 47.41 | [10.014, 107.314] [11.014, 89.214] |
| GRB110422A | 0.999 | 1.000 | 0.880 | 25.0 | 24.35 | [17.033, 27.133] [24.033, 26.133] |
| GRB071010B | 0.799 | 0.516 | 0.472 | 36.0 | 20.988 | [4.007, 39.208] [5.008, 38.208] |
| GRB051111 | 0.704 | 0.472 | 0.383 | 65.0 | 41.491 | [7.017, 77.317] [8.017, 75.317] |
| GRB050219A | 0.811 | 1.000 | 0.767 | 24.0 | 21.787 | [13.013, 25.113] [19.013, 24.113] |
| GRB150314A (no dim tail) | 0.989 | 1.000 | 0.260 | 11.0 | 11.149 | [4.004, 12.104] [11.004, 12.104] |
| GRB090510 | 0.652 | 0.432 | 0.402 | 5.69 | 2.70 | [0.064, 6.064] [0.164, 5.864] |

**Note.** In addition, we calculate the fraction of $T_{90,\mathrm{sim}}^{\mathrm{BAT}}$ values consistent with $T_{90,\mathrm{true}}^{\mathrm{BAT}}$ to within 3$\sigma$ and 1$\sigma$ limits for simulations using realistic parameter combinations (e.g., PCODE = 1, 0.522; NDETS = 20,000; BGDLVL = 95,000 counts s$^{-1}$). The majority of bursts have $f$(consistent, 1$\sigma$) ~25%–45%. GRB110422A is an exception, which we believe is due to the sharp shoulders seen on either side of the light curve. The Bayesian block analysis we use reports Gaussian uncertainties on duration measurements. In column 5, we report the $T_{90,\mathrm{true}}$ calculated for the 1 s resolution light curve used in the simulations. In column 6, we report the average $T_{90,\mathrm{sim}}$ calculated for all simulated light curves that were measurable by the Bayesian block algorithm. Lastly, in column 7, we report the 90% and 68% confidence interval (CI) for the sample.

light curve, $f$(consistent, 1$\sigma$), decreases steadily as the fluence of the synthetic light curve decreases (see Table 3). This trend is similar to what Kocevski & Petrosian (2013) found for FRED light curves folded with the CGRO/BATSE instrument response function. Yet, there are no clear trends between S/N and the measurement bias of $T_{90,\mathrm{sim}}^{\mathrm{BAT}}$ for real GRB light curves (see Table 3 and distributions in Appendix B); this is consistent with the results of Littlejohns et al. (2013).

Of the three observing conditions we evaluated, lowering the PCODE of the source (i.e., increasing the source angle from the detector bore sight) causes the largest bias on duration measurements (i.e., lower $T_{90,\mathrm{sim}}^{\mathrm{BAT}}$) and increases the uncertainty the most, followed by the average number of enabled detectors and background level. In Figure 7 we show that although the average $T_{90,\mathrm{sim}}^{\mathrm{BAT}}$ decreases as mask-weighted fluence decreases, distributions with similar fluence values show different structure depending on the observing conditions used in the respective simulations.

In Figure 22 we display plots similar to the bottom plots shown in Appendix B, but made using the measured mask-weighted fluence of the simulated light curves in the 15–350 keV band, $S_{\mathrm{sim}}^{\mathrm{BAT}}$. The mask-weighting process means the fluence has already been background subtracted and corrected for PCODE. We see that the fluence distributions are more tightly constrained than the $T_{90,\mathrm{sim}}$ distributions for each combination of observing conditions. This is most likely because it is typically the dim emission at the beginning and end of a burst that is easily lost into the noise, while the bright main-emission periods of a burst remain visible. So, while the $T_{90,\mathrm{sim}}$ may have large variations, the fluence remains relatively similar for a particular combination of observing conditions. However, even though the mask-weighting process aims to correct the $S_{\mathrm{sim}}^{\mathrm{BAT}}$ measurements for different observing conditions, the process cannot recover signal that is already lost; this is why we see that fluence decreases over one to two orders of magnitude with decreasing PCODE, depending on the GRB simulated. The fluence decreases by a factor of a few between NDETS = 30,000 and NDETS = 10,000. The complete figure set of fluence plots for all GRBs in our sample are available in the online journal; the remaining fluence plots are similar to the example shown in Figure 22.

### 5.2. Dim-tail Emission

Long, dim tails in a GRB light curve significantly increase the uncertainties of the duration measurements. Both GRB150314A (Figure 15) and GRB120119A (Figure 16) have long, dim tails (~100 s) in their light curves and are found to have $T_{90,\mathrm{sim}}^{\mathrm{BAT}}$ distributions which exhibit an arm extending much higher than the peak of the distribution. We simulated GRB150314A again but removed all signal outside of the interval T0−5 to T0+25 in order to remove the dim-tail emission (see bottom four plots of Figure 15). When the dim tail is removed, duration measurements are more tightly constrained around $T^{BAT}_{90,true}$ and the associated distributions do not show the extending arm.

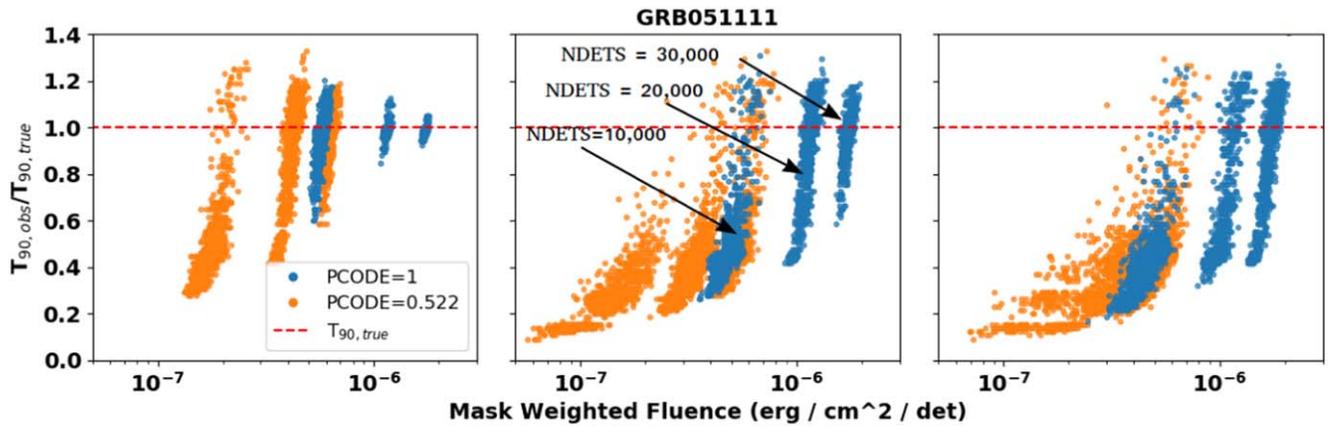

**Figure 7.** $T_{90}$ ratio values for GRB051111 separated by average background level (left: 1000 counts s$^{-1}$, center: 9500 counts s$^{-1}$, and right: 22,000 counts s$^{-1}$, respectively). There are three distinct groups for each PCODE value; these are due to the three numbers of enabled of detectors used during simulations. The groups are labeled in the center plot. We see that distributions with similar mask-weighted fluences may have more spread distribution shapes and behavior depending on the different observing conditions.

The lack of an extended arm in the $T^{BAT}_{90,sim}$ distributions of GRB150314A simulated without a tail implies that the Bayesian block method is unlikely to mistakenly measure noise fluctuations as dim-tail emission from the burst and, instead, when the Bayesian block algorithm does measure a dim tail, it is likely to be real. This may indicate a new method for measuring dim tails in light curves observed by Swift/BAT, i.e., when repeatedly measuring the $T_{90}$ of an observed light curve while applying noise fluctuations between each measurement, if the distribution of the measured $T_{90}$ shows an extended arm similar to those seen in distributions made for GRB150314A and GRB120119A, dim-tail emission may be present following the main emission.

Connaughton (2002) summed the background-subtracted signals from hundreds of CGRO/BATSE LGRBs and found gamma-ray tail emission lasting hundreds of seconds after their main emission period. They argued that this tail component is a common feature in LGRBs and independent of the duration and brightness of the prompt emission. This may imply that many more LGRBs observed with Swift/BAT have similar long, dim tails which, due to their low S/N, are often not measured (Connaughton 2002).

### 5.3. Collapsars Observed as SGRBs due to Observing Conditions

SGRBs are typically associated with compact-merger progenitor events, but there may be instances of collapsar events which are measured as SGRBs due to measurement bias arising from poor observing conditions. The template light curve generated from GRB160314A ($T_{90} \sim 8.7$; see Figure 9) is measured to have $T^{BAT}_{90,sim} < 2$ s in $\sim 1\%$ of all simulations; for these instances the progenitor system would be ambiguous if only the duration was considered. The duration distributions show a bimodal behavior (see Figure 14); some distributions are centered around the intrinsic duration while others display a strong instrument bias where most durations are measured to be $\sim 2.5$ s (see Figure 8). This may indicate that some collapsar events are measured as SGRBs due to instrument bias.

There is evidence that some SGRBs contain dim-tail emission that, if included as prompt emission, would increase the duration above 2 s and thus affect its classification as a SGRB. Connaughton (2002) applied the same background-subtraction method they applied to LGRBs to a sample of 100 SGRBs observed with CGRO/BATSE and found possible dim extended emission lasting for hundreds of seconds after the short, hard spike. Although, due to the low flux of the tail emission, the author does not claim if this is the same emission component found for the LGRB sample or, alternatively, if this emission is simply due to afterglow emission (Connaughton 2002). This latter option is supported by a study from Lazzati et al. (2001) who found X-ray afterglow in 76 short CGRO/BATSE GRBs by summing the four channel BATSE light curves. The emission peaked $\sim 30$ s after the short, hard spike of the prompt emission (Lazzati et al. 2001). There is theoretical support that afterglow emission from SGRBs will be separated from the prompt pulse by tens of seconds, while in LGRBs the afterglow will overlap the prompt emission (Sari & Piran 1999). The former option is supported by Dichiara et al. (2021), wherein the authors applied careful data reduction to eight Swift/BAT GRBs (all with $T_{90} < 2$ s and $z \geqslant 1$) and found that significant long, dim emission could be observed in four of the eight GRBs, which was not significant when using the standard Swift/BAT pipeline. This emission is found to be in excess of the X-ray afterglow emission and is therefore argued to be a prompt component. The authors show that the extended emission of GRB071227 would not be significant if the burst was at a higher redshift or was observed during a time of higher background (Dichiara et al. 2021).

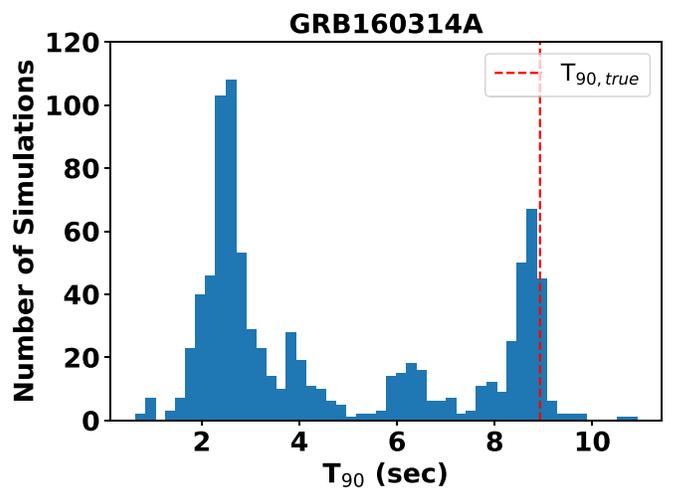

**Figure 8.** $T_{90,sim}$ distribution for GRB160314A using simulations made with PCODE = 0.522 and NDETS = 20,000. The intrinsic $T_{90}$ of the source is shown by the vertical red-dashed line ($\sim 8.7$ s). Many of the simulations have $T_{90}$ measurements $\sim 2.5$ s and some $<2$ s, suggesting a SGRB and implying a merger progenitor.

We included GRB090510 (Figure 21) in our sample of real GRB light curves. The light curve comprises a short, hard spike (∼0.3 s) and a soft tail (∼5 s). When only considering Fermi/GBM observations, the burst looks to be a typical short, hard GRB (Guiriec et al. 2009; Ackermann et al. 2010), but when including Swift/BAT observations, which display dim but significant excess emission following the short pulse, and considering the initial rise in optical emission, the distinction between short or long GRB becomes more complex (De Pasquale et al. 2010; Guiriec et al. 2010).

Of the ∼65.2% simulated light curves that were bright enough to obtain Bayesian block measurements, ∼55% had $T_{90,\text{sim}}^{\text{BAT}} < 2$ s and the remaining had $T_{90,\text{sim}}^{\text{BAT}} > 2$ s. Although we do not know how many SGRBs may exhibit similar behavior, it may be true that there are other GRBs which appear to only have a hard, short spike, but in reality include a dim, soft tail, though due to poor observing conditions the tail was not able to be measured. To truly remove the degeneracy between SGRBs and LGRBs, more information must be included in addition to the observed burst duration (e.g., burst environment, spectral information, and instrument condition upon observation).

## 6. Conclusion

We summarize our findings as follows:

1. The measured durations of simple-pulse GRB prompt-emission light curves observed with Swift/BAT are highly impacted by observing conditions. We see that, due to observing conditions, burst duration measurements commonly vary by ∼80% and fluence measurements can vary across two orders of magnitude.
2. As instrument sensitivity decreases (i) $T_{90}$ measurements become increasingly biased toward shorter durations, and (ii) the uncertainty of the measurements become larger.
3. We have shown that the PCODE has the strongest influence on measured duration and fluence, followed by the number of enabled detectors and then background level.
4. Light-curve shape has a strong influence on the accuracy of a $T_{90}$ measurement, even between similar S/N events. Consequently, it is difficult to quantify the trend between the S/N and the accuracy of the $T_{90}$.
5. For most GRBs in our sample, measured durations were consistent with their respective intrinsic durations only ∼25%–45% of the time.
6. Poor observing conditions, such as low PCODE, low NDETS, or high backgrounds, can cause source light curves with $T_{90,\text{true}} > 2$ s to be observed with $T_{90,\text{measured}} < 2$. This finding brings into question the viability of classifying all progenitor systems of SGRBs as merger events.

Based on our results, it is clear that the shape of a light curve highly impacts how accurate the observed duration is compared to the intrinsic burst duration. It will be worth investigating a larger sample of real light curves. Cosmological distance effects will be presented in a future paper.

We note that some of our parameter combinations are not likely to occur. In reality, the background stays near constant at ∼0.3 counts s$^{-1}$ per detector. A background of 1000 counts s$^{-1}$ will most likely not occur in reality, but it is a good reference point to compare to for this study. Additionally, a PCODE of 0.1 is uncommon. Although it is not expected currently, NDETS = 10,000 may be a reality in the future as the detector plane of Swift/BAT continues to degrade. In Table 3 we report the fraction of simulations with $T_{90}$ measurements consistent with the intrinsic light curve $T_{90}$, using only simulations with typical observing conditions (e.g., PCODE = 1, 0.522; NDETS = 20,000; BGDLVL = 95,000 counts s$^{-1}$). These fractions cannot be applied as correction factors to the Swift/BAT GRB population because the values of $f(\text{consistent},1\sigma)$ range from ∼7% to ∼88%, indicating that light-curve shape has a significant impact on the measured duration.

The durations of GRBs are important quantities that help infer progenitor systems and constrain physical processes of the bursts, yet the duration measurements are highly impacted by observing conditions. A proper understanding of the instrumental effects on measurements is necessary to correct for measurement bias. In any case, additional physical information must be used to accurately identify the progenitor system as a compact-binary merger or collapsar (e.g., burst environment, gravitational-wave measurements, and spectral behavior).

The authors would like to thank the anonymous reviewer for their very constructive and insightful comments. This work was in part funded by the Swift Guest Investigator program, Cycle 13 (80NSSC17K0336). All data used in this work were taken from the online Swift/BAT Gamma-Ray Burst Catalog.

## Appendix A
## Swift/BAT GRB Light Curves

Figure 9 displays the mask-weighted light curves for the eight real simple-pulse GRBs we used as light-curve templates in our simulations. Additional information on each GRB is presented in Table 2.

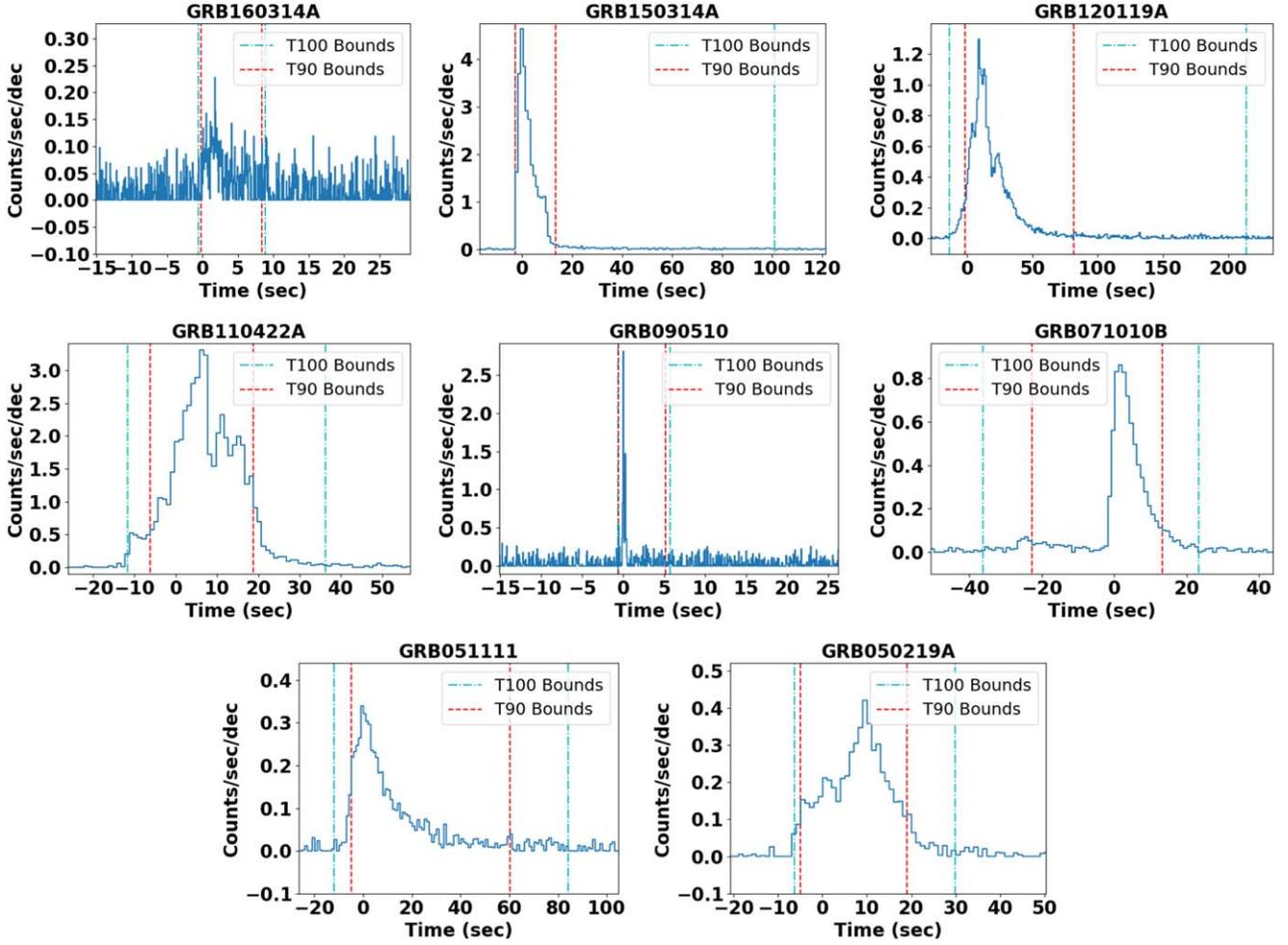

**Figure 9.** Our sample of simple-pulse-structured GRBs observed with Swift/BAT. These light curves are used as template light curves and are passed through the simulation pipeline. Information for each burst used to create these templates is located in Table 2.

## Appendix B
## $T_{90,\mathrm{sim}}$ Distributions

The measured $T_{90,\mathrm{sim}}^{\mathrm{BAT}}$ for FRED template light curves are displayed in Figures 10–13 and for real GRB light curves in Figures 14–21. In each figure, the top plot displays the template light curve used in the simulations (blue line) along with the $T_{90,\mathrm{true}}^{\mathrm{BAT}}$ of the template light curve (vertical red-dashed lines). Each horizontal orange bar is a single $T_{90,\mathrm{sim}}^{\mathrm{BAT}}$ measurement; the bars are ordered from longest to shortest duration (bottom to top). The number of orange bars is equal to the number of simulations able to be measured by the Bayesian block algorithm (e.g., $N_{\mathrm{bars}} = f(\mathrm{measurable}) * 27{,}000$). The bottom-left and right plots each display distributions of $T_{90,\mathrm{sim}}^{\mathrm{BAT}}$ at particular instrumental parameter combinations. The two plots contain $T_{90,\mathrm{sim}}^{\mathrm{BAT}}$ distributions made for simulations using different numbers of enabled detectors, i.e., simulations in the left plot used NDETS = 30,000 and the right plot 10,000. The distributions shown on each plot are shaded to indicate the average background level used during the simulation, i.e., dark blue $\sim$1000 c s$^{-1}$ and light blue $\sim$9500 c s$^{-1}$. Each plot displays the distributions at three different PCODE values indicated along the horizontal axis, i.e., PCODE = 0.101 ($\theta_{\mathrm{incident}} = 64°.286$), PCODE = 0.522 ($\theta_{\mathrm{incident}} = 44°.988$), and PCODE = 1.0 ($\theta_{\mathrm{incident}} = 0°.0$). The vertical axis on the left displays the ratio of the $T_{90,\mathrm{sim}}^{\mathrm{BAT}}$ to $T_{90,\mathrm{true}}^{\mathrm{BAT}}$; this can be also thought of as the ratio between one of the orange bars and the vertical red-dashed lines. The vertical axis on the right shows the duration measurement in seconds. When a distribution is tightly centered on $T_{90,\mathrm{sim}}^{\mathrm{BAT}}/T_{90,\mathrm{true}}^{\mathrm{BAT}} = 1$ (horizontal red-dashed line), the measured $T_{90,\mathrm{sim}}^{\mathrm{BAT}}$ of the simulated light curve accurately represents the $T_{90,\mathrm{true}}^{\mathrm{BAT}}$ of the template light curve.

For any combination of observing conditions where a distribution is not present in the figures below, then none of the simulations made for that combination of parameters resulted in a light curve that was bright enough for the Bayesian block algorithm to provide a duration measurement. In these cases, the GRBs would not have been observed.

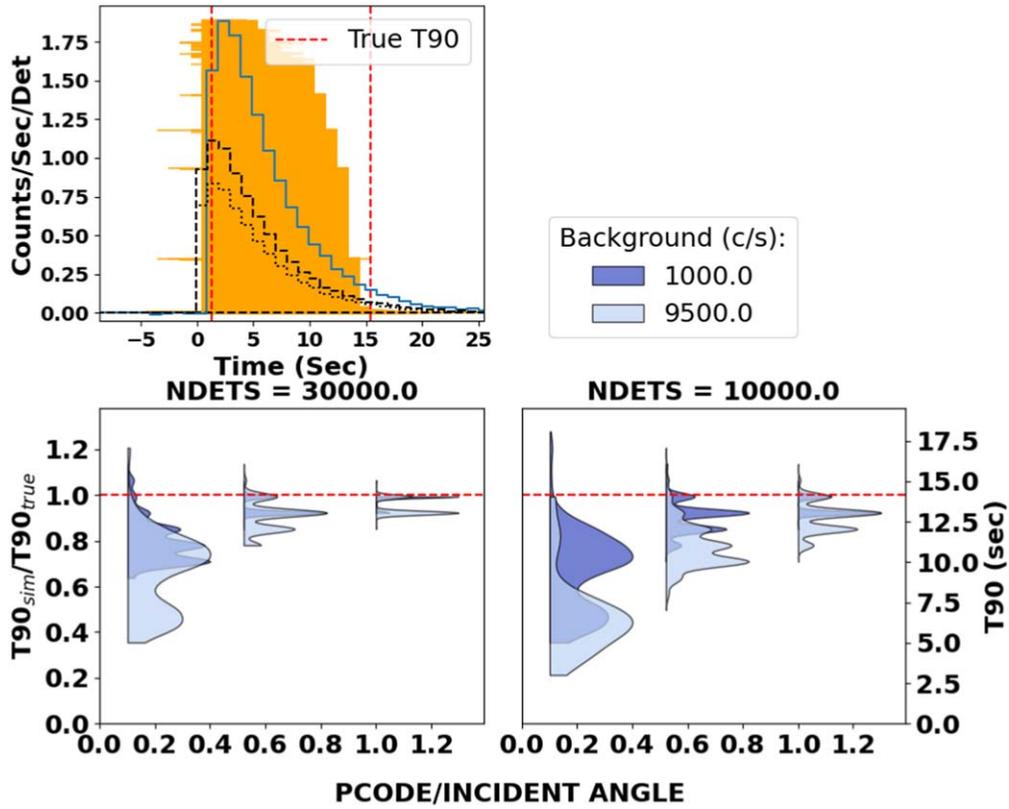

**Figure 10.** FRED1 Top: the template light curve (in blue) is displayed with two example simulated source light curves (both made with a background of ∼1000 c s$^{-1}$, but the dashed line uses PCODE = 1 and NDETS = 30,000, and the black dotted line uses PCODE = 0.522 and NDETS = 20,000). The vertical red-dashed lines show $T_{90,\text{true}}^{\text{BAT}}$. The horizontal orange bars show all measured $T_{90,\text{sim}}^{\text{BAT}}$; the number of orange bars is equal to the number of simulations able to be measured by the Bayesian block algorithm (e.g., $N_{\text{bars}} = f(\text{measurable}) * 27{,}000$). The two bottom plots display a series of distributions; each distribution is made up of the $T_{90,\text{sim}}^{\text{BAT}}$ measurements of $N = 1000$ synthetic observed light curves all simulated with the same observing conditions. Along the horizontal axis for the bottom plots is the partial coding fraction that the simulated observed light curves were simulated with (recall, PCODE = 1 denotes an on-axis observation, while PCODE = 0.1 denotes an incident angle of ∼ 60°). The different plots contain $T_{90}^{\text{BAT}}$ values measured for different numbers of enabled detectors (i.e., left: NDETS = 30,000 and right: NDETS = 10,000 ). There are two different shading colors for the displayed distributions, each referring to a different average background level (i.e., dark blue ∼1000 c s$^{-1}$ and light blue ∼9500 c s$^{-1}$). The ratio $T_{90,\text{sim}}^{\text{BAT}}/T_{90,\text{true}}^{\text{BAT}}$ is displayed on the left vertical axis and the actual measured durations in seconds on the right vertical axis.

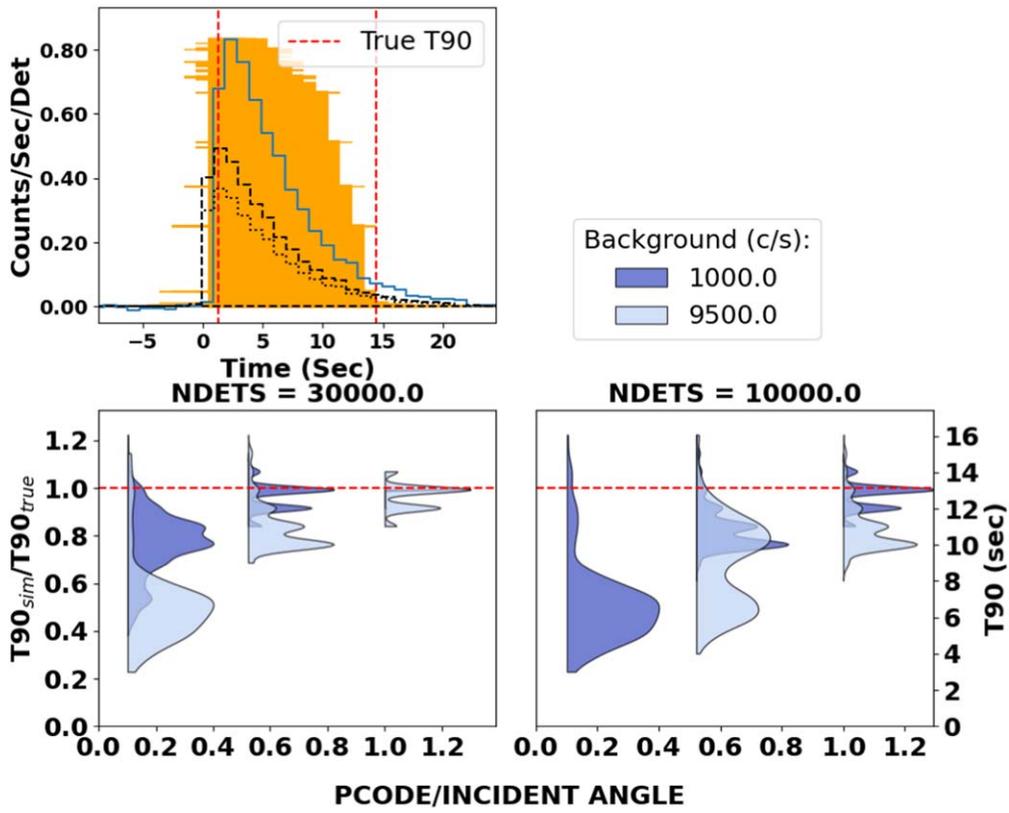

**Figure 11.** FRED2 (see plot descriptions in Figure 10).

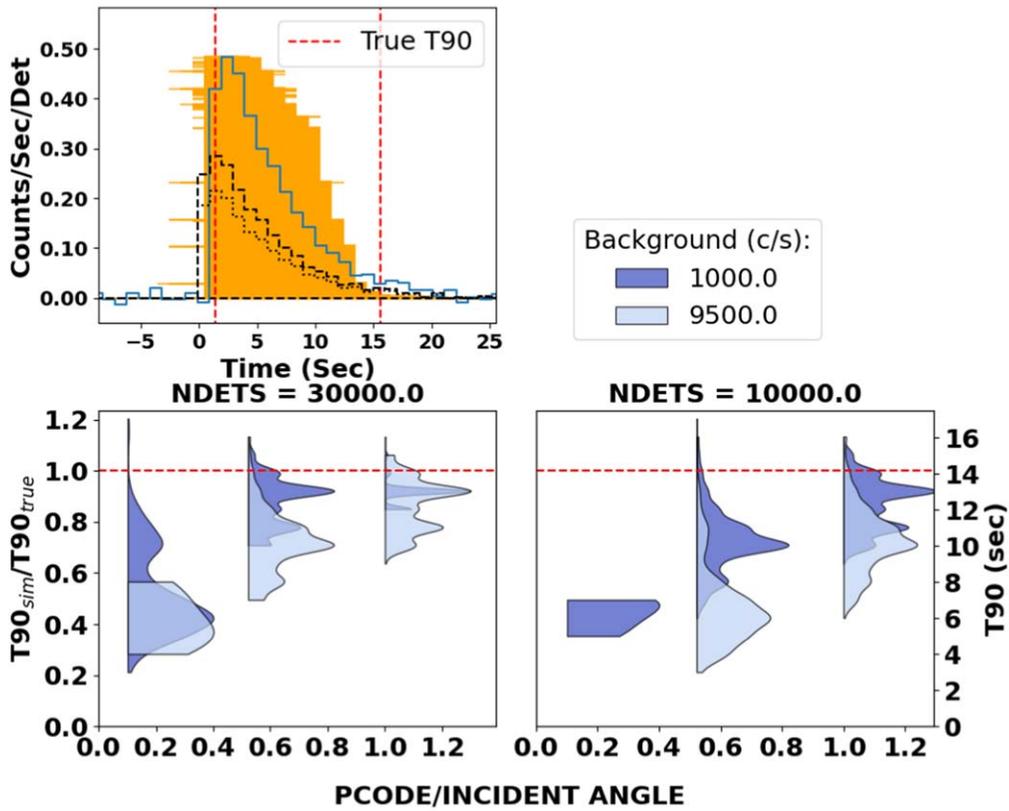

**Figure 12.** FRED3 (see plot descriptions in Figure 10).

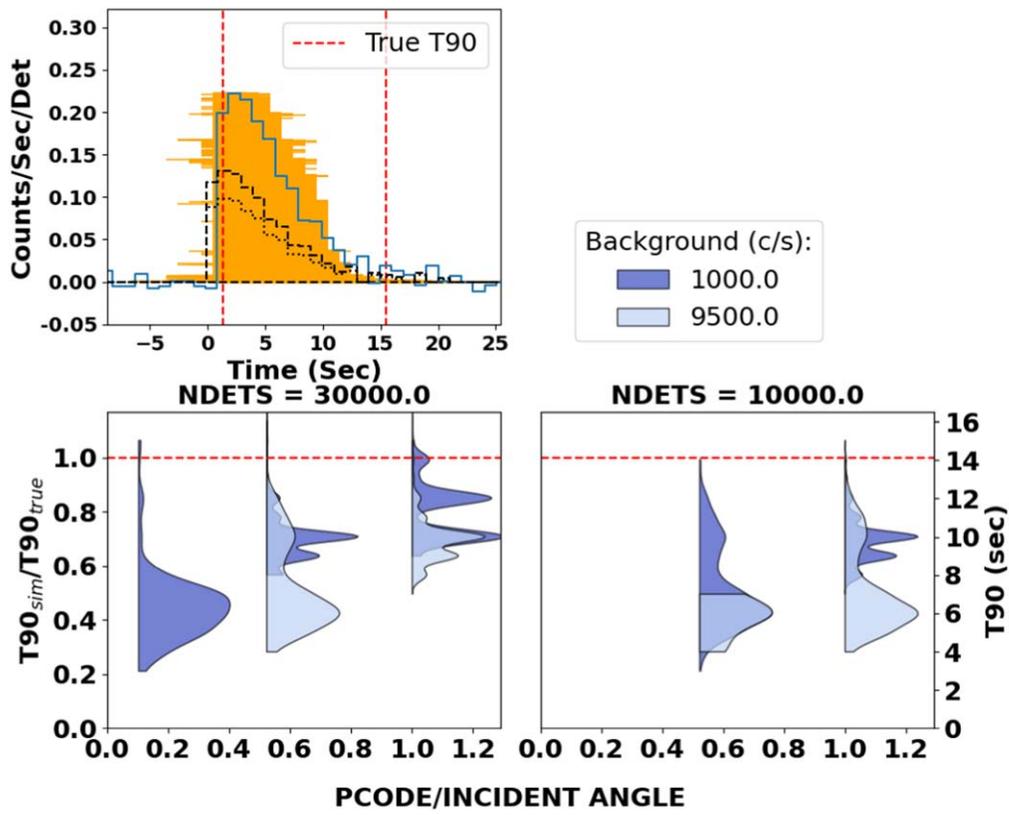

Figure 13. FRED4 (see plot descriptions in Figure 10).

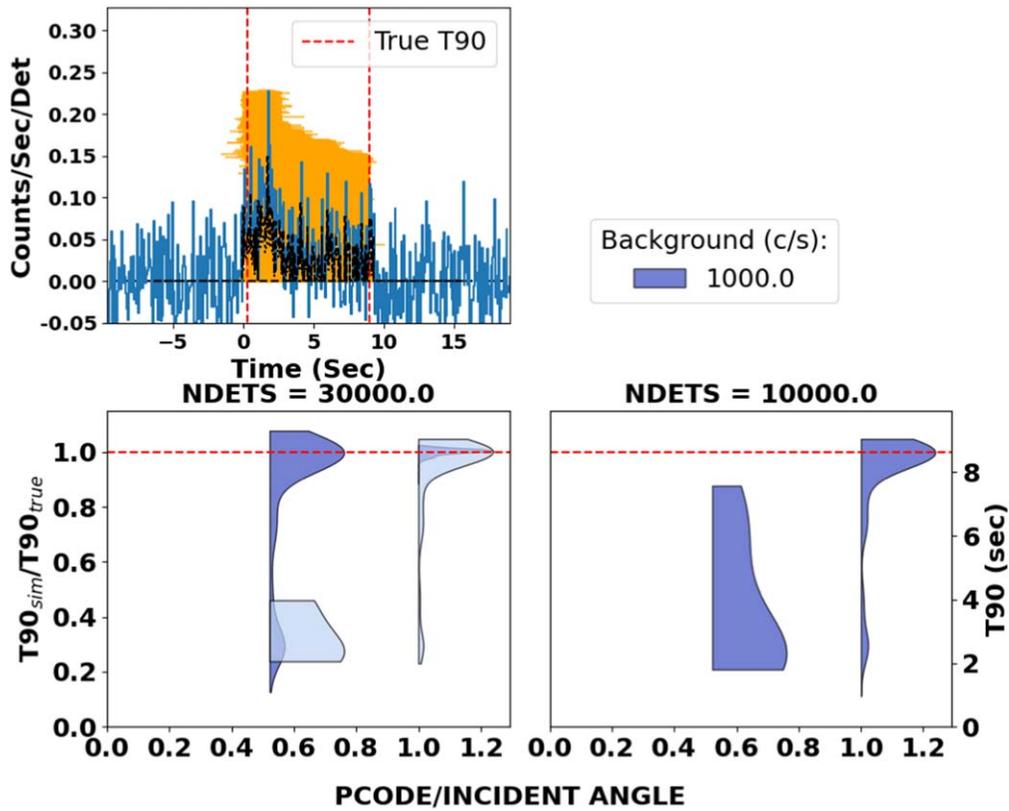

Figure 14. GRB160314A (see plot descriptions in Figure 10).

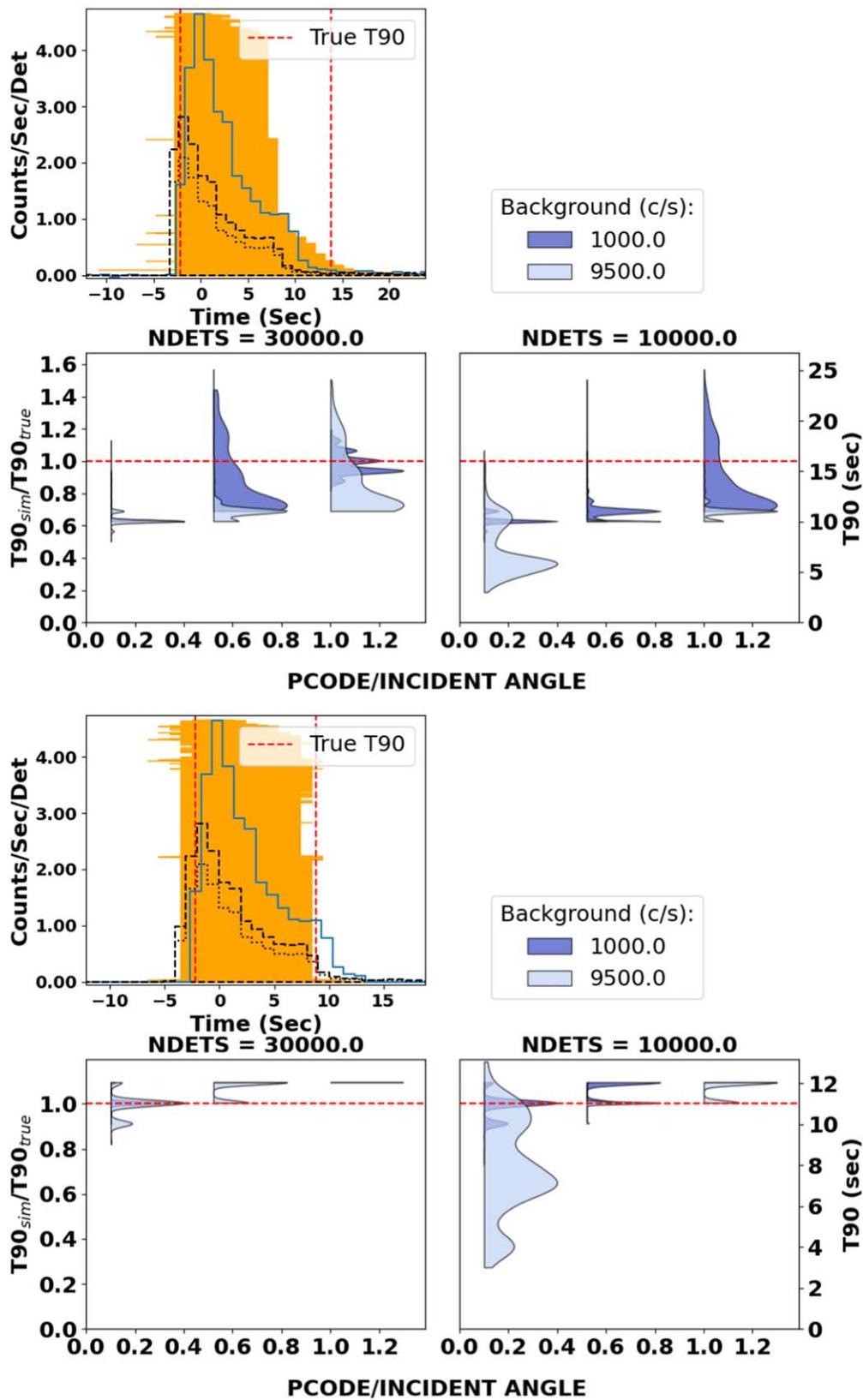

**Figure 15.** GRB150314A (see plot descriptions in Figure 10). The top three plots use a template light curve taken from the $T_{100}$ signal of GRB150314A. The bottom three plots again simulate GRB150314A as a template, but all emission outside of $T0 - \sim 5$ to $T0 + \sim 25$ s has been removed in order to remove the long, dim tail. Whereas the distributions which include the long, dim emission show arms extending to $T_{90,\mathrm{sim}} > T_{90,\mathrm{true}}$, the distributions created from simulations which exclude the long, dim emission are tightly constrained and do not display the same arm.

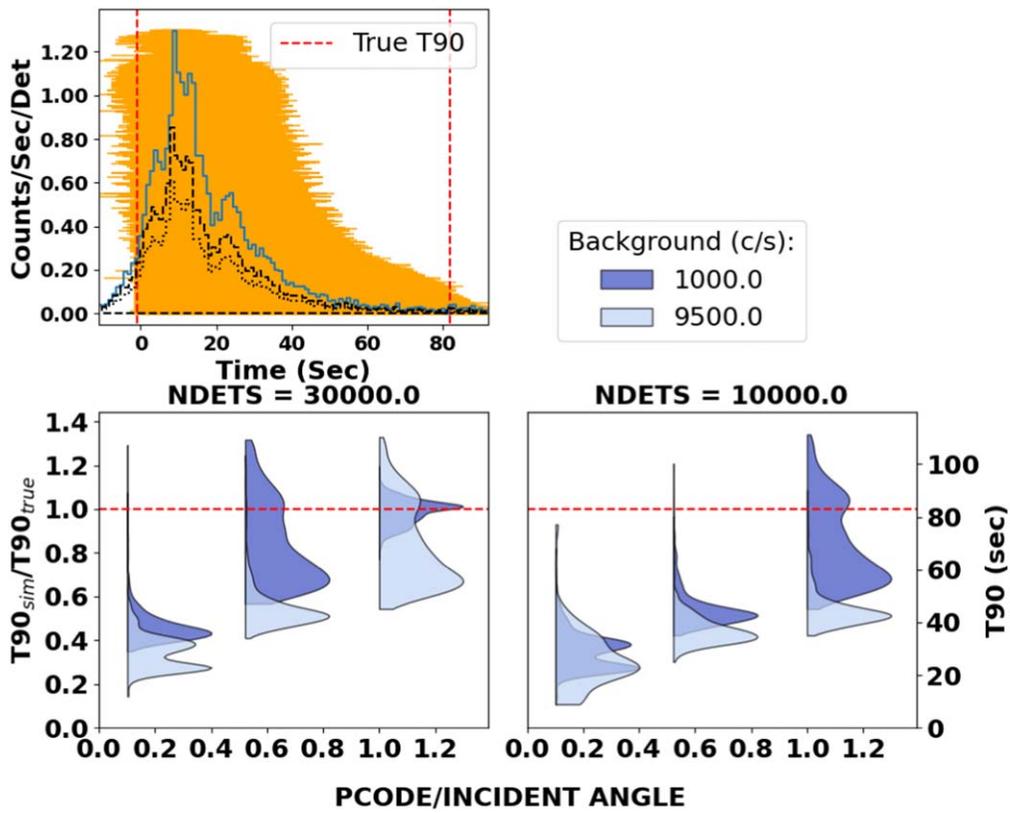

**Figure 16.** GRB120119A (see plot descriptions in Figure 10).

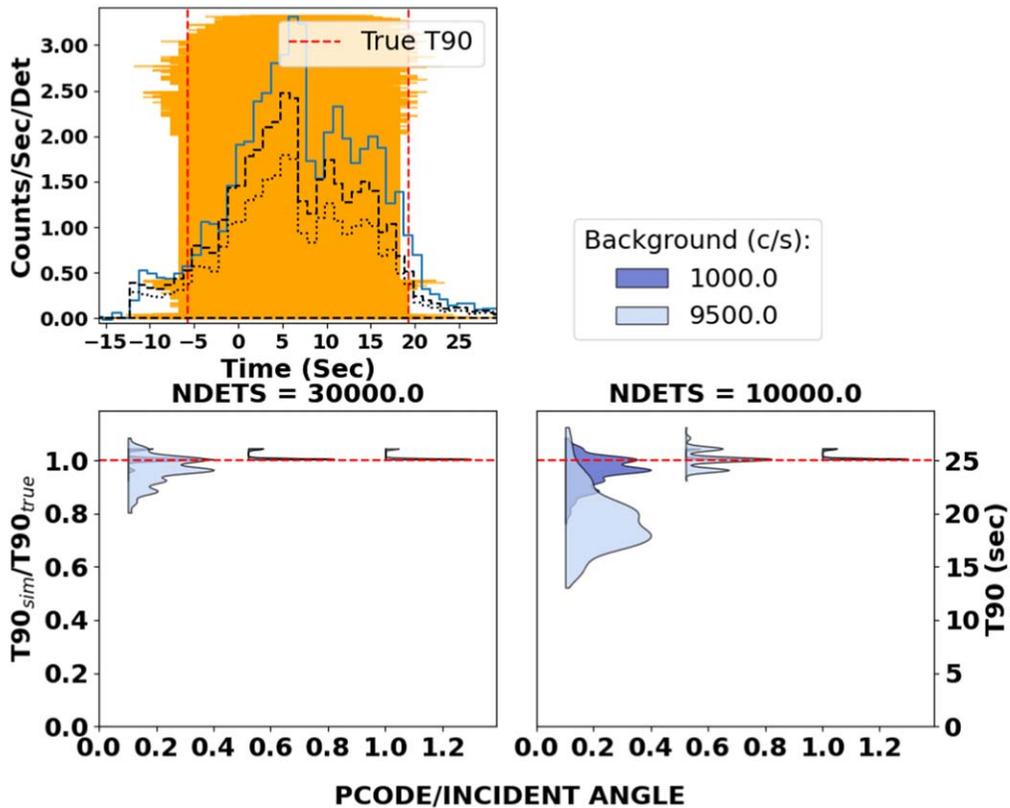

**Figure 17.** GRB110422A (see plot descriptions in Figure 10).

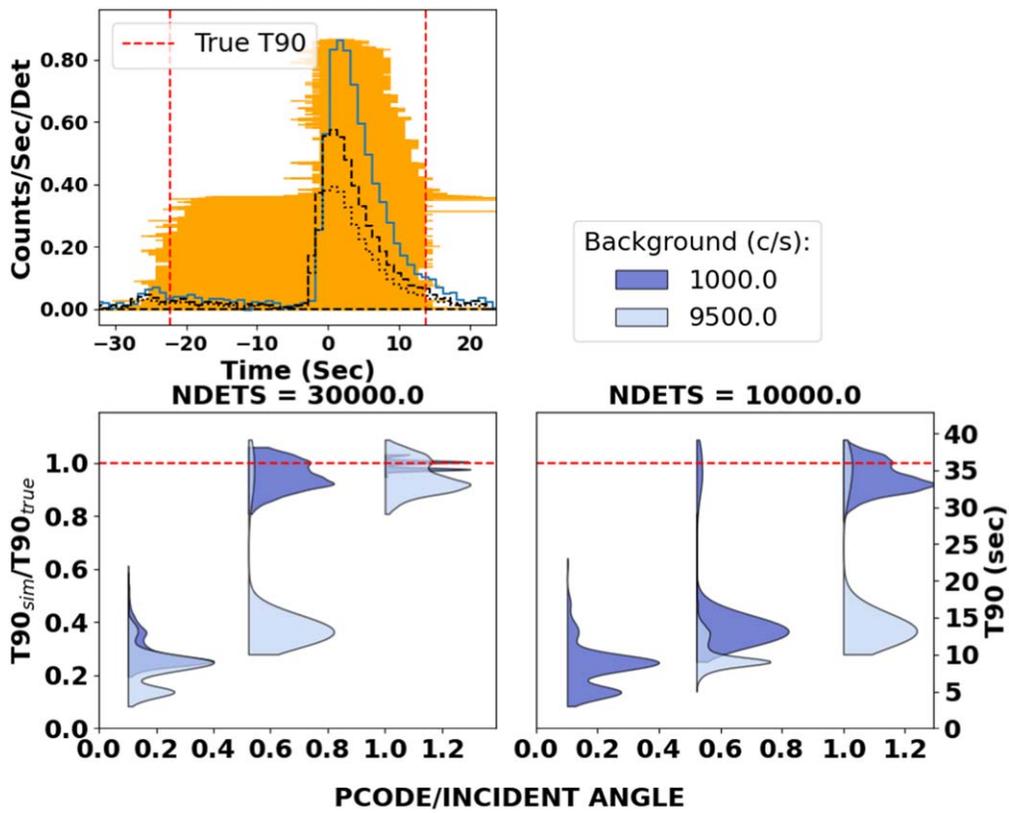

**Figure 18.** GRB071010B (see plot descriptions in Figure 10).

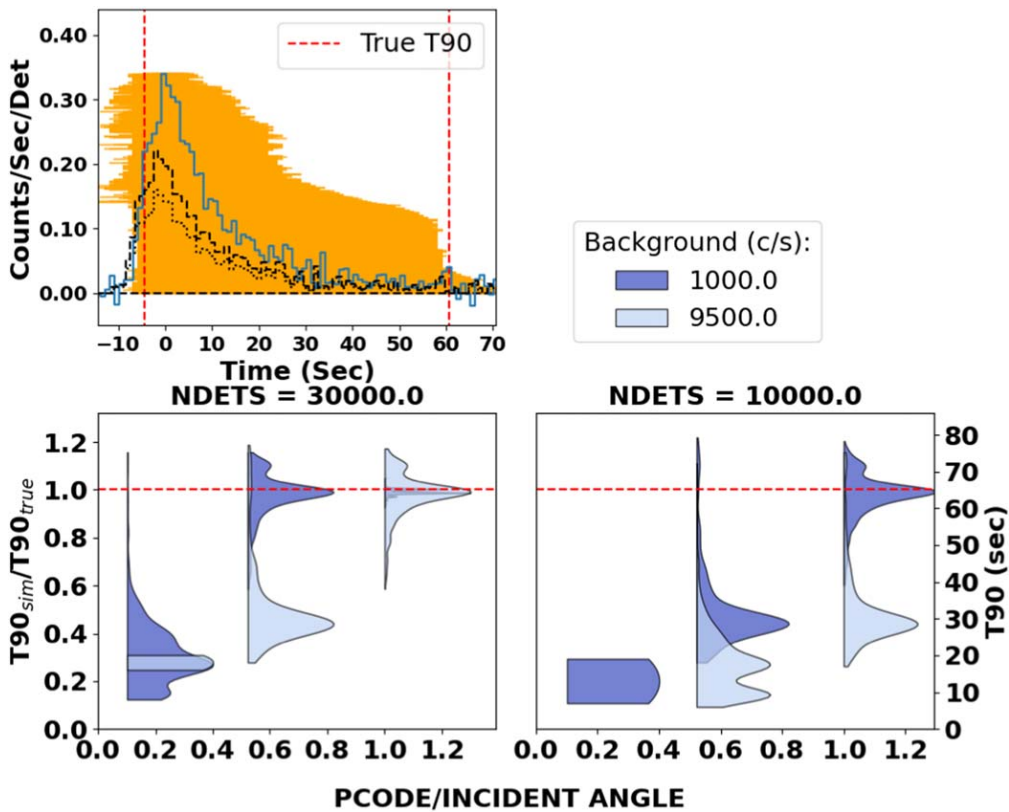

**Figure 19.** GRB051111 (see plot descriptions in Figure 10).

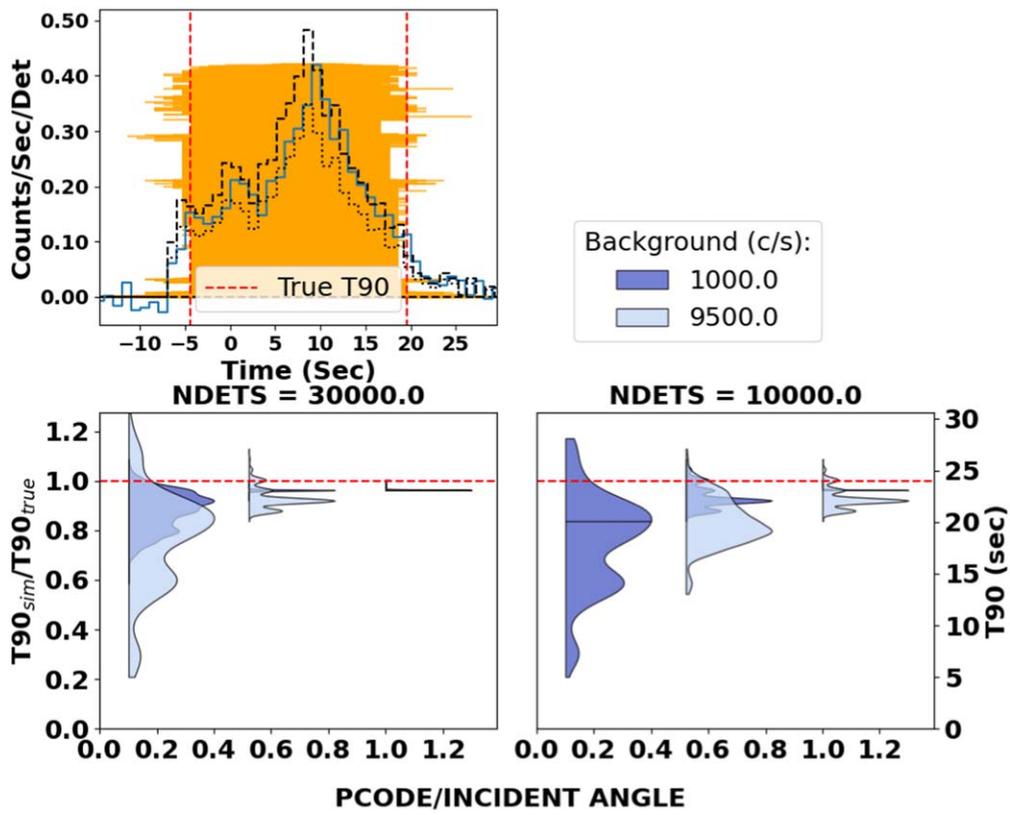

**Figure 20.** GRB050219A (see plot descriptions in Figure 10).

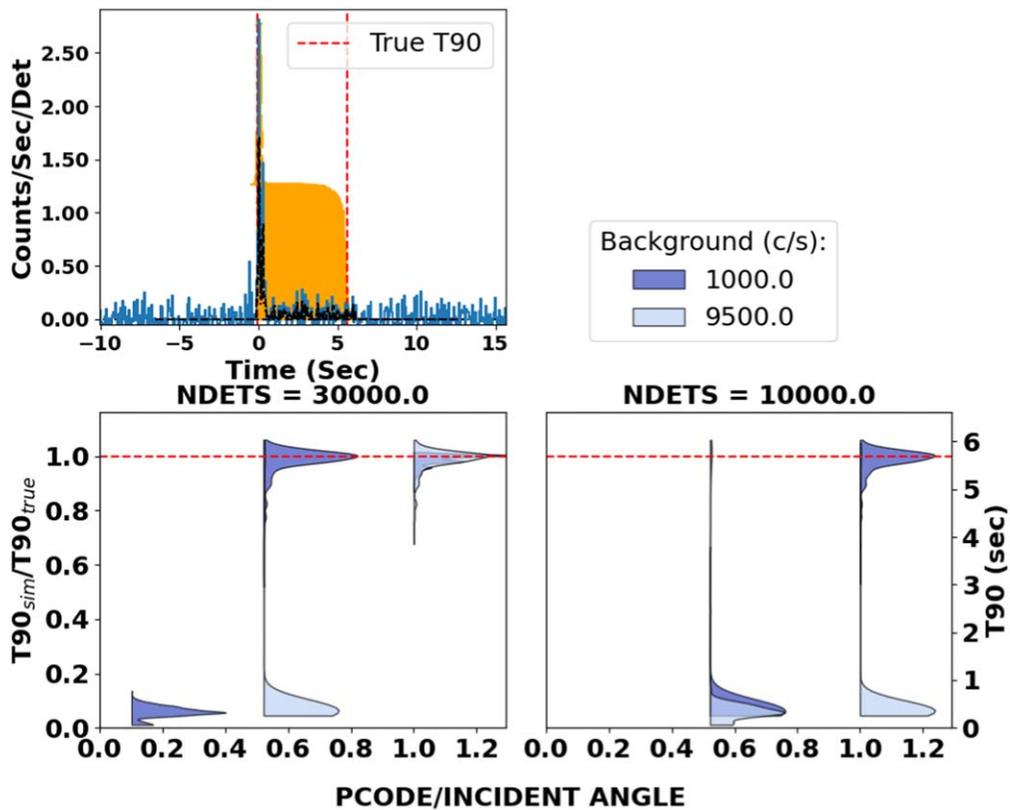

**Figure 21.** GRB090510 (see plot descriptions in Figure 10). This GRB was selected to demonstrate one possible case of a GRB with $T_{90,\mathrm{true}} > 2$ s being measured as a SGRB ($T_{90} < 2$ s) due to nonideal observing conditions. This effect becomes even worse when considering the rest frame of the burst, $z \sim 0.9$.

## Appendix C
## Fluence Distributions

The two plots in Figure 22 contain fluence distributions (the mask-weighted fluence in the 15–350 keV band) made from simulations using different numbers of enabled detectors, i.e., simulations in the left plot used NDETS = 30,000 and the right plot 10,000. The distributions shown on each plot are shaded to indicate the average background level used during the simulation, i.e., dark blue $\sim$1000 c s$^{-1}$ and purple $\sim$9500 c s$^{-1}$. Each plot displays the distributions at three different PCODE values indicated along the horizontal axis, i.e., PCODE = 0.101 ($\theta_{incident} = 64°.286$), PCODE = 0.522 ($\theta_{incident} = 44°.988$), and PCODE = 1.0 ($\theta_{incident} = 0°.0$). The vertical axis on the left displays the fluence value. The fluence plots for all GRBs in the sample can be viewed in the online journal.

For any combination of observing conditions where a distribution is not present in the figures below, then none of the simulations made for that combination of parameters resulted in a light curve that was bright enough for the Bayesian block algorithm to provide a duration measurement. In these cases, the GRBs would not have been observed.

## Appendix D
## Measurement Bias Dependency on Light-curve Phenomenology

### D.1. Synthetic FRED

For the synthetic FRED light curves (Figure 3), the duration measurements are increasingly biased toward shorter durations and measurement uncertainties increase directly as the brightness of the burst decreases (see Figures 10–13 and Table 3). In the case of FRED1 (with fluence $S(15–350$ keV$) = 1.44 \times 10^{-5}$ erg cm$^{-2}$; Figure 10), 91.4% of the simulations were measured above background and 27.8% were consistent with $T^{BAT}_{90,true}$ (at the 1$\sigma$ level). The dimmest synthetic FRED light curve, FRED4 ($S(15–350$ keV$) = 1.54 \times 10^{-6}$ erg cm$^{-2}$; Figure 13), was able to be measured in 57.1% of all simulations and only 10.1% remained consistent with $T^{BAT}_{90,true} = 14.14$ s (at the 1$\sigma$ level). This may imply that for real observations of dim FRED-like bursts, only $\sim$10% of the measured durations are representative of the intrinsic duration. It is important to note that a mask-weighted brightness of $\sim$0.2 counts s$^{-1}$ det$^{-1}$ (similar to FRED4; Figure 3) is not particularly uncommon for GRBs observed by Swift/BAT.

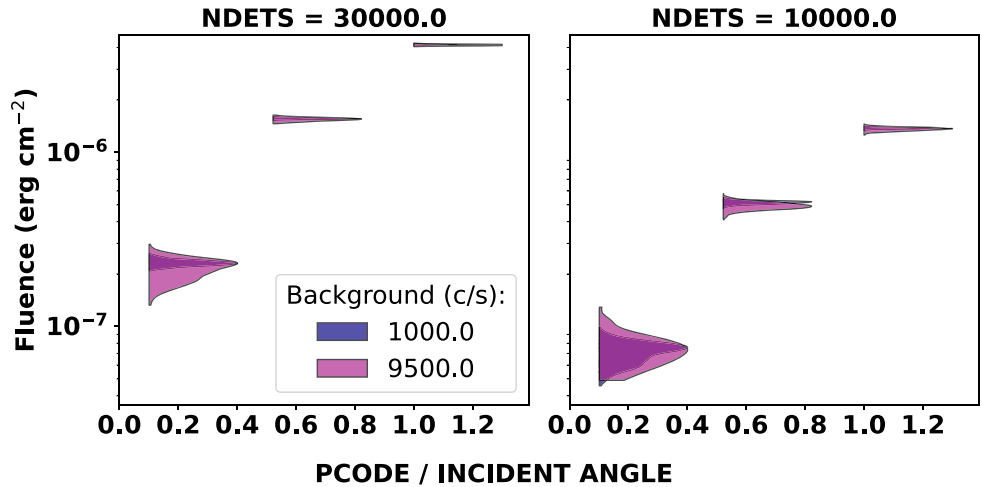

**Figure 22.** The two plots contain fluence distributions made from simulations of FRED1 using different numbers of enabled detectors, i.e., simulations in the left plot used NDETS = 30,000 and in the right plot 10,000. The fluence reported is the mask-weighted fluence measured in the 15–350 keV band. The distributions shown on each plot are shaded to indicate the average background level used during the simulation, i.e., dark blue $\sim$1000 c s$^{-1}$ and purple $\sim$9500 c s$^{-1}$. Each plot displays the distributions at three different PCODE values indicated along the horizontal axis, i.e., PCODE = 0.101 ($\theta_{incident} = 64°.286$), PCODE = 0.522 ($\theta_{incident} = 44°.988$), and PCODE = 1.0 ($\theta_{incident} = 0°.0$). The vertical axis on the left indicates the fluence. The complete figure set including the other light curves in our sample are available in the online journal.

(The complete figure set (12 images) is available.)

### D.2. Real GRBs: FRED-like

FRED1 and FRED4 have an order of magnitude difference in their fluences and are found to have $f(\text{consistent}, 1\sigma) = 0.278$ and 0.101, respectively. Yet for two light curves in our sample which have a similar order of magnitude difference in fluence, GRB120119A (Figure 16) and GRB050219A (Figure 20), the $f(\text{consistent}, 1\sigma)$ values are opposite to what we could expect, $f = 0.113$ and 0.563, respectively. Clearly, the shape of the light curve plays a role in the measurement bias of GRB durations.

A bimodal behavior is seen in the $T_{90,\text{sim}}^{\text{BAT}}$ distributions of GRB071010B (Figure 18). The light curve exhibits dim emission before the main pulse, which is often missed in nonideal instrument conditions, causing the $T_{90,\text{sim}}^{\text{BAT}}$ distributions to center around $\sim 0.4 * T_{90,\text{true}}^{\text{BAT}}$ in $\sim 66\%$ of simulations. A similar bimodal behavior is exhibited in GRB051111 (Figure 19) due to a dim tail following the main peak in the light curve.

### D.3. Real GRBs: Symmetric-like

From our sample, GRB110422A (Figure 17) and GRB050219A (Figure 20) both show the most consistent and narrow $T_{90,\text{sim}}^{\text{BAT}}$ distributions ($f(\text{consistent}, 1\sigma) = 0.820$, 0.563, respectively). We believe this accuracy is due to sharp shoulders, as can be seen in the light curves of each, which remain significant above the noise even in poor observing conditions. Interestingly, many of the $T_{90,\text{sim}}^{\text{BAT}}$ distributions for GRB050219A are very narrow, but centered around $T_{90}$ values shorter than the intrinsic duration.

### ORCID iDs

Michael Moss ● https://orcid.org/0000-0002-1103-7082
Amy Lien ● https://orcid.org/0000-0002-7851-9756
Sylvain Guiriec ● https://orcid.org/0000-0001-5780-8770
S. Bradley Cenko ● https://orcid.org/0000-0003-1673-970X
Takanori Sakamoto ● https://orcid.org/0000-0001-6276-6616